%% file: neurips_2026_revised.tex
\documentclass{article}

 \usepackage[main, final]{neurips_2026}

\usepackage[utf8]{inputenc} 
\usepackage[T1]{fontenc}    
\usepackage{hyperref}       
\usepackage{url}            
\usepackage{booktabs}       
\usepackage{amsfonts}       
\usepackage{nicefrac}       
\usepackage{microtype}      
\usepackage{xcolor}         
\usepackage{graphicx}
\usepackage{float}
\usepackage{subcaption}
\usepackage{multirow}
\usepackage[table]{xcolor}
\usepackage{amsmath}
\usepackage{amssymb}
\usepackage{algorithm}
\usepackage{algorithmic}
\usepackage{tabularx}
\usepackage{pifont}

\title{Virtual-Flow: Virtual Microphone-based Speech Enhancement via Semi-supervised Flow Matching}

\author{%
\textbf{Dongheon Lee}\textsuperscript{1\thanks{Work done during internship at Meta Reality Labs Research.}, 2}, \textbf{Ashutosh Pandey}\textsuperscript{1}, \textbf{Sanjeel Parekh}\textsuperscript{1}, \textbf{Zhaoheng Ni}\textsuperscript{1}, \\ \textbf{Daniel Wong}\textsuperscript{1}, \textbf{Jacob Donley}\textsuperscript{1}, \textbf{Buye Xu}\textsuperscript{1}, \textbf{Juan Azcarreta}\textsuperscript{1}\thanks{Corresponding Author} \\
  \textsuperscript{1}Meta Reality Labs Research, \hspace{1em} \textsuperscript{2}Korea Advanced Institute of Science and Technology (KAIST)\\
  \textsuperscript{1}Cambridge, Cambridgeshire, UK, \hspace{0.2em} Redmond, WA, USA, \hspace{1em} \textsuperscript{2}Daejeon, Republic of Korea \\
  \texttt{leedh0115@pknu.ac.kr, jazcarretao@meta.com}
}

\begin{document}

\maketitle
\vspace{-0.75em}

\begin{abstract}
  Neural network-based virtual microphone estimation (neural-VME) synthesizes microphone signals at unobserved spatial locations from a limited set of real microphone (RM) physical recordings. This enables high-resolution arrays that improve spatial processing for downstream tasks such as speech enhancement. However, existing methods rely on supervised training with ground-truth signals captured at virtual locations, requiring datasets that scale with the number of virtual microphones (VM) and leaving open the question of how to design optimal large-array configurations. We propose  \textit{Virtual-Flow}, a flow-matching framework that implicitly generates VM signals conditioned on real recordings, without requiring ground-truth VM signals. We introduce a semi-supervised training strategy based on pseudo-VM targets constructed from time-delayed superpositions of real recordings. Virtual-Flow shows improved magnitude estimation in high-frequency bands, yielding richer spatial cues for beamforming. Experiments demonstrate that the proposed Virtual-Flow, combined with reflow distillation, outperforms existing neural-VME baselines in VM-based beamforming and achieves competitive end-to-end speech enhancement performance with low computational cost.
\end{abstract}
\vspace{-0.75em}

\section{Introduction} Multichannel speech enhancement is the task of removing noise and reverberation from multichannel noisy reverberant speech to estimate the clean speech of the target speaker \cite{wang2018all, lee2024deft, quan2024spatialnet, pandey2025ultra}. Increasing the number of available microphones raises spatial sampling density, enabling finer-grained spatial cues that help distinguish competing sources \cite{wang2020derev, ochiai2020beam, benesty2008microphone, VanVeen:1988}. In practice, however, consumer devices such as AR glasses \cite{donley2021easycom}, earbuds \cite{bae2024speech}, and hearing aids \cite{fedorov2020tinylstms} are equipped with only a handful of microphones, fundamentally limiting the spatial information available for enhancement. 

Neural network-based virtual microphone estimation (neural-VME) has been proposed to address this constraint \cite{ochiai2021neural}. Neural-VME estimates virtual microphone (VM) signals as synthesized observations at unobserved spatial positions, given real microphone (RM) signals measured by physical sensors. Predicting VM signals encourages the model to learn rich spatial representations, which can then be exploited by VM-based beamforming (VM-BF) \cite{ochiai2021neural, segawa2022neural}, which utilizes beamforming as the back-end of the speech enhancement framework.

However, most existing neural-VME methods focus on estimating a single VM signal \cite{ochiai2021neural, segawa2022neural, segawa2024neural, qiu2024transformer, qiao2025neural, chen2025uniarray}, even though leveraging a large number of VM signals can significantly enhance spatial resolution \cite{farmani2026beamforming, pan2025enhancing}. Scaling to many VM channels in a supervised setting is challenging for two reasons: it demands larger datasets with ground-truth recordings at each virtual position, and it requires a fixed array configuration that may not be optimal for the downstream task.

On the modeling side, previous works employ conventional discriminative speech enhancement architectures for neural-VME \cite{segawa2024neural, qiu2024transformer, qiao2025neural, pan2025enhancing}, despite the inherent differences between the two tasks. The recently proposed Spatial-Magnifier \cite{lee2026spatial} achieves high performance by reformulating neural-VME as a generative task using a generative adversarial network (GAN) \cite{goodfellow2014generative}. However, generating signals across a large and variable set of virtual positions requires both stable training over diverse target distributions and efficient inference. Flow matching \cite{lipman2023flow} offers both: stable optimization via continuous-time probability paths and fast inference through reflow distillation \cite{liu2023flow}.

Consequently, two critical gaps remain: (1) developing scalable methods to exploit a \textbf{large number of VM signals} for VM-BF, and (2) establishing a \textbf{semi-supervised neural-VME framework} that optimizes generated channels for downstream enhancement without ground-truth VM signals. Here, semi-supervised neural-VME requires no ground-truth VM signals. Clean reference and excitation speech are used only during training for enhancement supervision and pseudo-target construction, respectively; inference uses only RM signals.

In this work, we propose \textit{Virtual-Flow}, a flow-matching framework that addresses both gaps by generating numerous VM signals without ground-truth VM signals. Our key contributions are:

1. \textbf{Virtual-Flow}: We design a flow-matching model that can estimate a large number of VM signals, achieving state-of-the-art VM-BF performance. Through reflow distillation \cite{lipman2023flow, liu2023flow}, the model enables single-step inference with low computational cost.

2. \textbf{Semi-supervised neural-VME method}: We propose a pseudo-VM target generation strategy based on time-delayed superpositions of RM signals, providing approximate supervision in delay space without requiring real VM recordings. This enables effective semi-supervised flow matching training through end-to-end optimization with the downstream beamforming objective.

Even with only four RM signals, our method substantially closes the gap to neural beamforming with sixteen physical microphones. Prior work has estimated more than ten VM signals using acoustics-informed neural networks in controlled array settings \cite{zhao2024circular}. Our contribution is to estimate twelve VM signals from four RMs for speech enhancement under noisy, reverberant conditions using a flow-matching framework. Extensive experiments demonstrate that Virtual-Flow outperforms neural-VME baselines with and without ground-truth VM supervision while requiring lower computational cost. We further show that meaningful performance gains persist even without pseudo targets, using only the beamforming objective. Additionally, we explore extensions to end-to-end VM-based speech enhancement (VM-SE), as well as array-agnostic and real-time variants for practical deployment.

\section{Background and Related Work}
\subsection{Neural Network-based Virtual Microphone Estimation (Neural-VME)}
Ideally, $M$-channel noisy reverberant speech $\mathbf{y} \in \mathbb{R}^{M \times N}$, with $N$ time-domain samples, can be decomposed in terms of the direct speech $\mathbf{x} \in \mathbb{R}^{M \times N}$ through the following mathematical modeling:
\begin{align}
    \mathbf{y} &= \mathbf{x} + \mathbf{x}_{rev} + \mathbf{n}
   \\&= \mathbf{o} * \mathbf{h}^d + \mathbf{x}_{rev} + \mathbf{n},\label{eq:modeling}
\end{align}

where $\mathbf{o} \in \mathbb{R}^{1\times N}, \mathbf{x}_{rev}, \mathbf{n} \in \mathbb{R}^{M \times N}$, and $\mathbf{h}^d$ denote the excitation speech, the reverberant component of the target speech, the additive background noise recorded by the microphones, and the impulse response of the direct-path signal, respectively. $*$ denotes the convolution operation.

However, in this study, we assume a practical scenario where only $M_r$ microphones are available on actual physical devices $(M_r < M)$. Multichannel speech enhancement (MC-SE) aims to estimate the direct-path target speech at a reference microphone $\mathbf{x}_{ref} \in \mathbb{R}^{1 \times N}$, modeled as:
\begin{align}
    \hat{\mathbf{x}}^{se}_{ref}&=\text{MC-SE}(\mathbf{y}_r),
\end{align}

where $\mathbf{y}_{r}\in \mathbb{R}^{M_r\times N}$ denotes measured RM signals. Neural-VME estimates the VM signals $\mathbf{y}_{v}\in \mathbb{R}^{M_v\times N}$ given RM signals using a neural network:
\begin{align}
\begin{aligned}
\hat{\mathbf{y}}_{v} &= \text{Neural-VME}(\mathbf{y}_{r}),
\end{aligned}
\end{align}
where $M_v$ denotes the number of VM channels.

Following the VM-BF framework \cite{ochiai2021neural}, an augmented signal $\hat{\mathbf{y}} = [\mathbf{y}_{r}, \hat{\mathbf{y}}_{v}] \in \mathbb{R}^{M \times N}$ (where $M = M_r + M_v$) is constructed to compute spatial covariance matrices (SCM). In our framework, the enhanced signal $\hat{\mathbf{x}}^{se}_{ref}$ from a pre-trained MC-SE model is approximated as a reference to derive the adaptive beamforming coefficients, followed by beamforming (BF) with an augmented signal,
\begin{align}
\begin{aligned}
\hat{\mathbf{x}}_{ref}^{bf}=\text{BF}(\hat{\mathbf{x}}^{se}_{ref}, \hat{\mathbf{y}}).
\end{aligned}
\end{align}
The beamforming output $\hat{\mathbf{x}}_{ref}^{bf} \in \mathbb{R}^{1\times N}$ serves as the estimated target signal of the reference channel. The adaptive beamforming coefficients can be analytically determined in closed form using classical beamforming criteria, such as the multichannel Wiener filter (MCWF) \cite{benesty2008microphone} or the minimum variance distortionless response (MVDR) beamformer \cite{souden2009optimal}. By jointly optimizing neural-VME and the beamforming objective, the model learns to generate VM signals that effectively increase the numerical rank of the SCM from $M_r$ to $M$, thereby improving beamforming performance \cite{segawa2024neural}. 
The detailed procedure of adaptive beamforming is described in Appendix~\ref{app:beamformers}.

\subsection{Spatial Audio Representation Learning (SARL)}
Since conventional neural-VME utilizes the augmented noisy signal $\hat{\mathbf{y}}$ only for neural beamforming, it cannot be directly employed for end-to-end speech enhancement. To address this limitation, Lee \textit{et al.} ~\cite{lee2026spatial} proposed Spatial Audio Representation Learning (SARL), which optimizes the enhanced signal $\hat{\mathbf{x}}^{se}_{ref}$ by augmenting real microphone (RM) observations with virtual spatial information. In this work, we leverage the SARL-signal (SARL-S) framework, where the neural-VME model estimates explicit VM signals to augment the input $\hat{\mathbf{y}} = [\mathbf{y}_{r}, \hat{\mathbf{y}}_{v}]$ of the MC-SE model:
\begin{align}
\hat{\mathbf{x}}_{ref}^{se_{\hat{\mathbf{y}}}} &= \text{MC-SE}(\hat{\mathbf{y}}),
\end{align}
which enables the downstream MC-SE model to leverage enhanced spatial cues across the expanded array geometry. The output from the enhanced MC-SE model can either be fed into an adaptive beamformer to perform VM-BF or be used directly as the final output for VM-SE.

\section{Method: Virtual-Flow with Semi-supervised Learning}
\textbf{Supervision setting.} In this paper, semi-supervised refers to VM-based enhancement trained with clean reference speech $\mathbf{x}_{ref}$ for the downstream objective, but without ground-truth VM signals $\mathbf{y}_v$. VME therefore lacks direct VM-target supervision, while the complete enhancement system retains clean-speech supervision. At inference, only the observed RM signals $\mathbf{y}_r$ are required. We apply this distinction consistently to the evaluated baseline configurations as well as Virtual-Flow. Table~\ref{tab:signal_availability} distinguishes training and inference resources. 
\begin{table}[!htbp]
\vspace{-1em}
\centering
\caption{Signal availability in semi-supervised learning of Virtual-Flow.}
\label{tab:signal_availability}
\setlength{\tabcolsep}{4pt}
\scalebox{0.85}{
\begin{tabularx}{\linewidth}{lccX}
\toprule
Signal & Training & Inference & Role \\
\midrule
RM signals $\mathbf{y}_r$ & \ding{52} & \ding{52} & Model input \\
Excitation $\mathbf{o}$ & \ding{52} & \ding{56} & Delay estimation for pseudo-target construction \\
Clean reference $\mathbf{x}_{ref}$ & \ding{52} & \ding{56} & BF-loss target \\
VM signals $\mathbf{y}_v$ & \ding{56} & \ding{56} & No ground-truth VM supervision \\
\bottomrule
\end{tabularx}}
\vspace{-1em}
\end{table}

\begin{figure}[!t]
\centering
\vspace{-1em}
\includegraphics[width=5.5in]{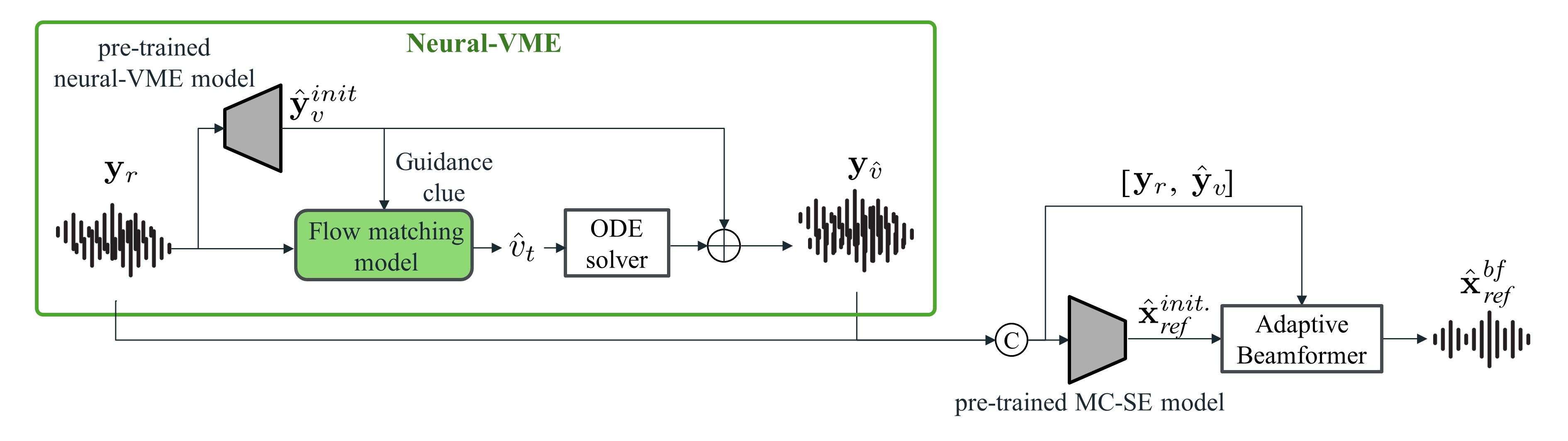}
\vspace{-2em}
\caption{Proposed framework for virtual microphone-based beamforming (VM-BF). A pre-trained neural-VME model provides a guidance cue to the flow matching model, enabling precise estimation of the vector field. The VM-noisy speech is reconstructed during both training and inference, thereby enhancing the MC-SE model and improving downstream neural beamforming performance.}
\label{fig:framework}
\end{figure}

\begin{algorithm}[!b]
\caption{VM-BF training algorithm using our flow matching model}\label{alg:flow_matching}
\begin{algorithmic}
\STATE $\textbf{Requires:}\;\text{pre-trained MC-SE, neural-VME model}$
\STATE $\textbf{Inputs:}\;\text{RM signals}\;\mathbf{y}_r,\;\text{clean reference}\;\mathbf{x}_{ref},\;\text{training target}\;\mathbf{y}_v^\star=\tilde{\mathbf{y}}_v$ from Algorithm~\ref{alg:unsupervised}.
\STATE $\hat{\mathbf{y}}_v^{init.} = \text{Neural-VME}(\mathbf{y}_r)$
\hfill{\textcolor{magenta}{\textit{initialized neural-VME output}}}
\STATE \textbf{for} $s \sim \mathcal{U}(0, 1)$ \textbf{or} $s=0$ \textbf{do}
\STATE \hspace{0.5cm} $\mathbf{z}=\tilde{\mathbf{y}}_r + \sigma \boldsymbol{\epsilon}, \quad \boldsymbol{\epsilon} \sim \mathcal{N}(\mathbf{0},\mathbf{I})$
\STATE \hspace{0.5cm} $\phi_s(\mathbf{z})=s\mathbf{y}_v^\star + (1-s)\mathbf{z} $
\STATE \hspace{0.5cm} $\psi_s=\mathbf{y}_v^\star-\mathbf{z}$
\STATE \textbf{end for}
\STATE$\mathcal{L}_{\mathrm{CFM}}
= \mathbb{E}_{s}\Big[\Big\|
\mathrm{STFT}\!\big(u_s(\phi_s(\mathbf{z}) \mid \hat{\mathbf{y}}_{v}^{init.})\big)
- \mathrm{STFT}(\mathbf{y}_v^\star - \mathbf{z})
\Big\|_1\Big]$ \hfill{\textcolor{magenta}{\textit{conditional flow matching loss}}}
\STATE $\hat{\mathbf{y}}_v = \tilde{\mathbf{y}}_r + \int_{0}^{1} u_s(\phi_s(\mathbf{z}) | \hat{\mathbf{y}}_v^{init}) ds$  \hfill{\textcolor{magenta}{\textit{vector field integration via ODE solver}}}
\STATE $\hat{\mathbf{x}}_{ref}^{init.} = \text{MC-SE}([\mathbf{y}_r, \hat{\mathbf{y}}_{v}])$
\STATE $\hat{\mathbf{x}}_{ref}^{bf}=\text{BF}(\hat{\mathbf{x}}_{ref}^{init.}, [\mathbf{y}_r, \hat{\mathbf{y}}_{v}])$ \hfill{\textcolor{magenta}{\textit{VM-BF (Details in Appendix~\ref{app:beamformers})}}}
\STATE $\mathcal{L}_{\mathrm{total}}=w_{vme}\log_{10}\mathcal{L}_{\mathrm{CFM}}+w_{bf}\mathcal{L}_{\mathrm{SNR}}(\hat{\mathbf{x}}_{ref}^{bf},\mathbf{x}_{ref})$
\end{algorithmic}
\end{algorithm}

\subsection{Conditional Flow Matching model}
Our Virtual-Flow model preserves the view of neural-VME as a generation task, while enabling high-quality generation and fast inference via flow matching. The task definition for our flow matching-based neural-VME is given by
\begin{align}
\begin{aligned}
\hat{\mathbf{y}}_{v} = f_{\theta}\!\left(\mathbf{y}_r \mid \hat{\mathbf{y}}_{v}^{init.}\right),
\end{aligned}
\end{align}
where $f_{\theta}, \hat{\mathbf{y}}_{v}^{init.}$, and $\hat{\mathbf{y}}_{v}$ denote the flow matching model, initialized neural-VME output, and estimated neural-VME output using the flow matching model, respectively. Here, $\mathbf{y}_v$ denotes the true noisy VM signals, while $\hat{\mathbf{y}}_v$ denotes the model estimate; they are distinct and need not coincide. We denote the training endpoint by $\mathbf{y}_v^\star$: it equals $\mathbf{y}_v$ for supervised neural-VME and the pseudo-target $\tilde{\mathbf{y}}_v$ for semi-supervised neural-VME. Since the input $\mathbf{y}_r$ must match the channel dimensionality ($M_v$) of the target $\mathbf{y}_v^\star$, we pad the number of RM channels ($M_r$) by appending $M_v - M_r$ copies of the noisy observation at the reference channel as $\tilde{\mathbf{y}}_r = [\mathbf{y}_r;\mathbf{y}_{ref};\cdots;\mathbf{y}_{ref}] \in \mathbb{R}^{M_v \times N}$, assuming $M_v > M_r$. We formulate the flow matching problem using a time-dependent flow function $\phi_s(\mathbf{z})$ defined by a linear interpolation between $\mathbf{y}_v^\star$ and an auxiliary variable $\mathbf{z}$ over $s \in [0,1]$:
\begin{align}
\begin{aligned}
\phi_s(\mathbf{z}) = s\mathbf{y}_v^\star + (1-s)\mathbf{z},
\end{aligned}
\end{align}
For each fixed $s\in[0,1]$, $\phi_s : \mathbb{R}^{M_v \times N} \rightarrow \mathbb{R}^{M_v \times N}$; the subscript $s$ indexes the family of flow maps. Here, $s$ denotes a continuous time step for the flow process, which should be distinguished from the discrete temporal index of the audio signal. Using the re-parameterization technique, we express $\mathbf{z}$ as $\mathbf{z} = \tilde{\mathbf{y}}_r + \sigma \boldsymbol{\epsilon}$ 
where $\boldsymbol{\epsilon} \sim \mathcal{N}(\mathbf{0}, \mathbf{I})$ and $\sigma$ is a scalar controlling the perturbation magnitude. $\mathbf{I}$ denotes identity matrix. The probability path is
\begin{align}
\begin{aligned}
p_s(\mathbf{z}) =
\mathcal{N}\!\left(\mathbf{z}\,\middle|\, s\mathbf{y}_v^\star + (1-s)\tilde{\mathbf{y}}_r,\ (1-s)^2\sigma^2\mathbf{I}\right),
\end{aligned}
\end{align}
and at $s=0$ it reduces to
\begin{align}
\begin{aligned}
p_0(\mathbf{z}) = \mathcal{N}\!\left(\mathbf{z}\,\middle|\, \tilde{\mathbf{y}}_r,\ \sigma^2\mathbf{I}\right).
\end{aligned}
\end{align}
We can derive the vector field $\psi_s$ from the following ordinary differential equation (ODE):
\begin{align}
\begin{aligned}
\frac{d}{ds}\phi_s(\mathbf{z}) = \psi_s(\phi_s(\mathbf{z})).
\end{aligned}
\end{align}
The training target vector field along the path is
\begin{align}
\begin{aligned}
\psi_s(\phi_s(\mathbf{z})) = \mathbf{y}_v^\star - \mathbf{z},
\end{aligned}
\end{align}
while the model predicts a conditional vector field $u_s(\phi_s(\mathbf{z}) \mid \hat{\mathbf{y}}_{v}^{init.})$. We train the model using a conditional flow matching (CFM) loss defined as an $\ell_1$ distance in the STFT domain:
\begin{align}
\mathcal{L}_{\mathrm{CFM}}
= \mathbb{E}_{s}\Big[\Big\|
\mathrm{STFT}\!\big(u_s(\phi_s(\mathbf{z}) \mid \hat{\mathbf{y}}_{v}^{init.})\big)
- \mathrm{STFT}(\mathbf{y}_v^\star - \mathbf{z})
\Big\|_1\Big].
\end{align}
The relationship between neural-VME and flow matching is further detailed in Appendix~\ref{app:flow_matching}.

\subsection{Model Architecture}
\textbf{Overall framework} The overall VM-BF framework using SARL-S is shown in Figure~\ref{fig:framework}. Virtual-Flow estimates the VM signals $\hat{\mathbf{y}}_v$, which are concatenated with the RM signals to form the augmented input $\hat{\mathbf{y}} = [\mathbf{y}_r, \hat{\mathbf{y}}_v]$. This augmented signal is then provided to a pre-trained MC-SE model that controls an adaptive neural beamformer. The adaptive beamformer employs linear filtering to mitigate non-linear distortions caused by MC-SE neural network processing, a widely adopted strategy in prior work \cite{yamaoka2019cnn, segawa2022neural, segawa2024neural, wang2024unsupervised, qiu2024transformer, hattori2025evaluation, qiao2025neural, farmani2026beamforming, wang2026fixed}.

\begin{figure*}[!t]
\centering
\includegraphics[width=5.2in]{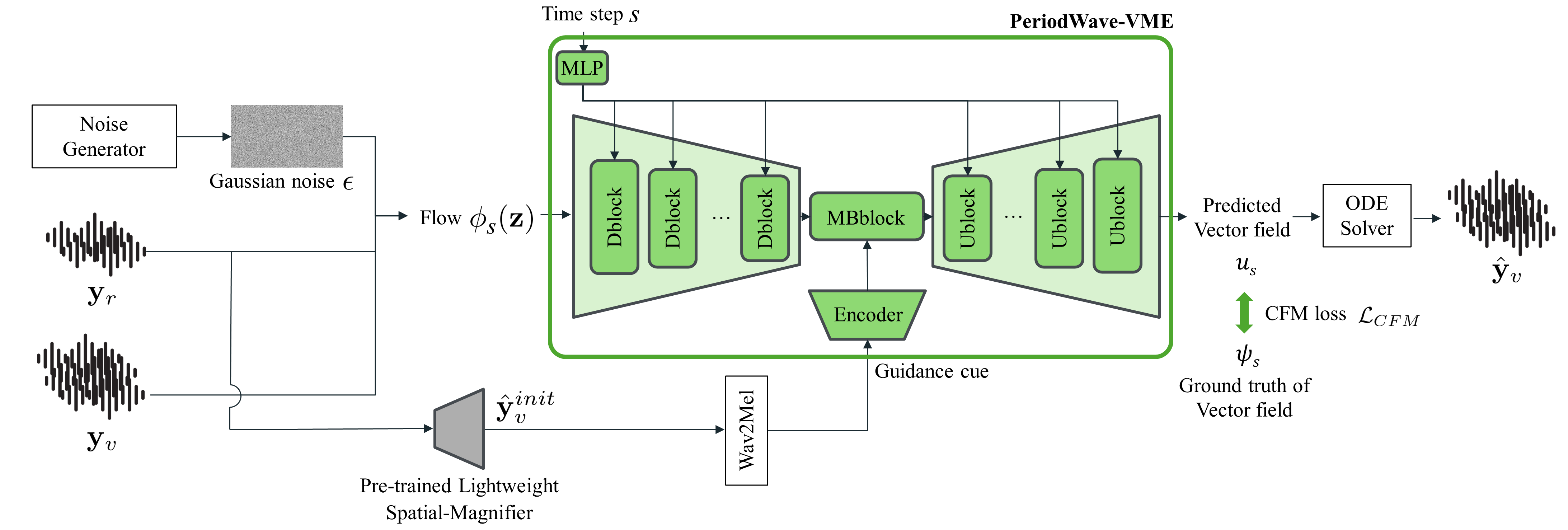}
\vspace{-0.5em}
\caption{Overall architecture of Virtual-Flow. The RM signals are first passed through the pre-trained neural-VME model (lightweight Spatial-Magnifier) to obtain a guidance cue, which the flow matching model (PeriodWave-VME) then uses to estimate the vector field.}
\label{fig:model}
\vspace{-1.5em}
\end{figure*}

\textbf{Flow matching training with reflow} In a conventional flow matching pipeline, the model is trained to predict the vector field at a random time step $s \in [0,1]$, and generates the final sample at inference by integrating the predicted field with an ODE solver. In our framework, however, neural-VME serves as an intermediate module whose ultimate objective is not signal reconstruction itself, but improved downstream enhancement. Accordingly, we run the ODE solver during training to obtain $\hat{\mathbf{y}}_v$ and propagate gradients through the full VM-BF pipeline. We further adopt reflow \cite{liu2023flow}, which first trains the model with random time steps $0 \leq s \leq 1$, and then re-trains the model to predict the straight vector field ($s=0$). This enables VM signals generation with a single function evaluation (NFE$=1$), yielding fast inference without significant quality loss. Algorithm~\ref{alg:flow_matching} provides pseudo-code for the training algorithm for the proposed flow matching-based VM-BF.

\textbf{Virtual-Flow architecture} Virtual-Flow comprises two branches, illustrated in Figure~\ref{fig:model}: a pre-trained discriminative model (lightweight Spatial-Magnifier) and a flow matching model (PeriodWave-VME). The lightweight Spatial-Magnifier produces initial VM signal estimates that serve as a guidance cue for the flow matching model. Lightweight Spatial-Magnifier is a simplified variant of \cite{lee2026spatial} by compressing the channel dimension, removing the selection module and dynamic channel allocation (DCA), and replacing all convolution operations with group convolutions. Furthermore, to accelerate the training process, we trained the model exclusively as a discriminative model. In the semi-supervised setting, this guidance model is pre-trained using only the beamforming loss with clean reference speech, without ground-truth VM targets. At both training and inference, the guidance cue $\hat{\mathbf{y}}_v^{init.}$ is computed solely from $\mathbf{y}_r$; clean reference and excitation speech are never guidance inputs.

PeriodWave-VME is derived from PeriodWave \cite{lee2025periodwave}, which we adopted as our backbone architecture due to its capacity for lightweight and high-fidelity audio synthesis. The input waveform is chunked into periodic segments and processed in parallel by a parameter-shared U-Net. After the down (D)-blocks, latent features are compressed to a bottleneck representation and fused with the guidance cue from the lightweight Spatial-Magnifier via skip connections in the multi-band (MB)-block. The up (U)-blocks then restore the features to the vector-field dimensionality. Unlike the original PeriodWave, we provide the guidance cue at both training and inference to avoid a domain gap between the two stages. The output vector field is integrated via an ODE solver to produce the final VM signals $\hat{\mathbf{y}}_v$. Furthermore, we extend our model to be real-time and array-agnostic in Appendix \ref{app:rt_aga}.

\subsection{Semi-supervised Learning Method}\label{sec:semi_supervised}
\begin{figure*}[!t]
\centering
\includegraphics[width=5.5in]{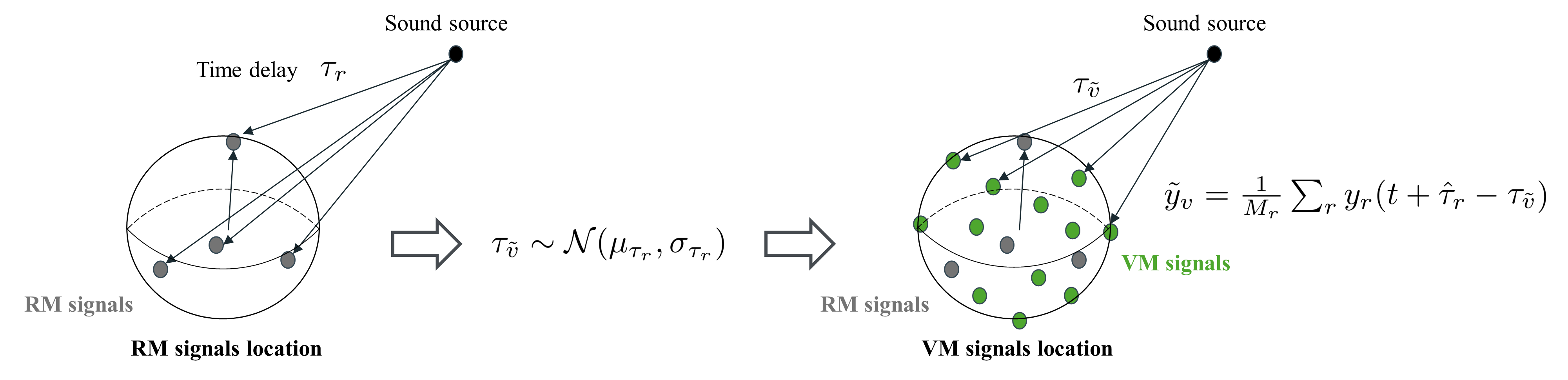}
\vspace{-1.5em}
\caption{Overview of semi-supervised flow matching using pseudo targets. The pseudo-VM targets are calculated by superposing time-shifted RM signals. The virtual time delays are selected as evenly spaced points derived from the mean and standard deviation of the estimated RM time delays.}
\label{fig:unsupervised}
\vspace{-1em}
\end{figure*}

\textbf{Semi-supervised flow matching with pseudo targets} We use the term semi-supervised because clean reference speech supervises downstream enhancement, while no ground-truth VM signals are used to train neural-VME. The excitation speech $\mathbf{o}$ is used only during training to estimate delays and construct the pseudo-VM target $\tilde{\mathbf{y}}_v$; it is not itself the VM target in the CFM loss. The clean reference $\mathbf{x}_{ref}$ supervises the BF loss. Specifically, we first estimate the time delay $\tau_r$ between the excitation speech $o[t]$ (defined in Eq.~\ref{eq:modeling}) and each RM signal $y_r[t]$ by maximizing their cross-correlation:
\begin{align}
\begin{aligned}
\hat{\tau}_r = \arg\max_{\tau_r} \sum_{t} o[t]y_r[t+\tau_r].
\end{aligned}
\end{align}
From the estimated delays $\{\hat{\tau}_r\}$, we compute the mean $\mu_{\tau_r}$ and standard deviation $\sigma_{\tau_r}$, then select $M_v$ virtual delays on an equidistant grid:
\begin{align}
\begin{aligned}
\tau_{\tilde{v},k}=\mu_{\tau_r}+\left(\frac{2(k-1)}{M_v-1}-1\right)\left\lfloor\frac{M_v}{4}\right\rfloor\sigma_{\tau_r},\quad k=1,\ldots,M_v. \tau_{\tilde{v}} \sim \mathcal{N}\!\left(\mu_{\tau_r}, \sigma_{\tau_r}\right).
\end{aligned}
\end{align}
The deterministic grid spans from $\mu_{\tau_r} - \lfloor{\frac{M_v}{4}}\rfloor\sigma_{\tau_r}$ to $\mu_{\tau_r} + \lfloor{\frac{M_v}{4}}\rfloor\sigma_{\tau_r}$. This provides diverse delay-conditioned training targets for the VM channels. We draw $M_v$ virtual delays and generate the pseudo target of the VM noisy signal by a shifted superposition of RM signals:
\begin{align}
\begin{aligned}
\tilde{y}_v[t] = \frac{1}{M_r}\sum_{r=1}^{M_r} y_r\!\left[t+\hat{\tau}_r-\tau_{\tilde{v}}\right].
\end{aligned}
\end{align}
The pseudo-target therefore supplies an approximate acoustic prior for the target speaker's delayed direct-path component. Reverberation and noises need not align in this way; the pseudo-target is not an exact reconstruction of noisy speech at a physical VM location. The BF loss further adapts the generated channels for downstream enhancement. Unlike position-based interpolation restricted to the segment between RMs, this construction selects targets in delay space without imposing that geometric constraint. A delay alone does not uniquely specify a 3D location, so this flexibility does not establish accurate synthesis at arbitrary physical positions. Figure~\ref{fig:unsupervised} illustrates the pseudo-target construction.
Algorithm~\ref{alg:unsupervised} provides pseudo-code for the proposed semi-supervised flow matching.
\begin{algorithm}
\caption{Semi-supervised learning method using pseudo-VM targets}\label{alg:unsupervised}
\begin{algorithmic}
\STATE $\textbf{Requires:}\;\text{RM noisy speech}\;\mathbf{y}_r, \text{excitation speech}\;\mathbf{o}$
\STATE $\hat{\tau}_r=\underset{\tau_r}{\operatorname{argmax}}{\sum_{t}o[t] y_r[t+\tau_r]}$ \hfill{\textcolor{magenta}{\textit{estimation of RM time delays}}}
\STATE $\{\tau_{\tilde{v}}\} = \text{Linspace}(\mu_{\tau_r} - \lfloor\frac{M_v}{4}\rfloor\sigma_{\tau_r}, \mu_{\tau_r} + \lfloor\frac{M_v}{4}\rfloor\sigma_{\tau_r}, M_v)$ \hfill{\textcolor{magenta}{\textit{sampling of VM time delays}}}
\STATE $\tilde{y}_v[t]=\frac{1}{M_r}\sum_{r=1}^{M_r}y_r[t+\hat{\tau}_{r}-\tau_{\tilde{v}}]$ \hfill{\textcolor{magenta}{\textit{pseudo-VM targets}}}
\STATE \textbf{Return:}\;$\tilde{\mathbf{y}}_v$
\end{algorithmic}
\end{algorithm}

\textbf{Semi-supervised SARL} SARL-S requires a pre-trained MC-SE model whose input consists of $M = M_r + M_v$ microphone channels. However, under the semi-supervised setting, pre-training such an MC-SE model is not feasible because the corresponding $M$-channel observations are unavailable. To address this issue, we propose semi-supervised SARL-S. We first pre-train an MC-SE model using only the $M_r$ microphone channels. Given that most MC-SE models typically adopt an encoder-separator-decoder architecture \cite{luo2018tasnet, luo2019conv, wang2023tf, quan2024spatialnet, lee2025deft}, the encoder is trained from scratch with convolution kernels that accept $M$ channels during neural-VME training. Meanwhile, the separator and decoder are initialized from the pre-trained $M_r$-channel MC-SE model and fine-tuned.

\textbf{Loss function}
The total loss function is defined as a weighted sum of the neural-VME loss and the beamforming (BF) loss \cite{segawa2024neural}. Given the absence of ground-truth VM signals for semi-supervised neural-VME, we employ a semi-supervised flow matching loss using pseudo target $\tilde{\mathbf{y}}_v$. In contrast, since the clean speech target is available, the BF loss is formulated as the signal-to-noise ratio (SNR) loss between the estimated speech and the direct clean speech. To align the scale with the SNR of the BF loss, we applied a base-10 logarithm to the CFM loss.
\begin{align}
\begin{aligned}
\mathcal{L}_{\mathrm{total}}
= w_{vme}\log_{10}\mathcal{L}_{\mathrm{CFM}}(\hat{\mathbf{y}}_{v},\tilde{\mathbf{y}}_v)
+ w_{bf}\mathcal{L}_{\mathrm{SNR}}(\hat{\mathbf{x}}_{ref}^{bf},\mathbf{x}_{ref}),
\end{aligned}
\end{align}
Additionally, the loss function for the supervised learning is formulated as described in Appendix \ref{app:loss}.

\subsection{Inference}\label{sec:inference}
At inference, the trained system takes only the noisy RM signals $\mathbf{y}_r$ as external input. Excitation speech $\mathbf{o}$, clean reference speech $\mathbf{x}_{ref}$, ground-truth VM signals $\mathbf{y}_v$, and pseudo-VM targets $\tilde{\mathbf{y}}_v$ are not required. At inference, we set $\boldsymbol{\epsilon}=\mathbf{0}$ and initialize the flow with the padded RM signals $\tilde{\mathbf{y}}_r$. For fixed model parameters, inference is deterministic and follows four steps:
\begingroup
\setlength{\leftmargini}{0.5\leftmargini} \setlength{\parskip}{0pt}
\begin{enumerate}
\item \textbf{Guidance generation.} The pre-trained lightweight Spatial-Magnifier maps $\mathbf{y}_r$ to the initial VM estimates $\hat{\mathbf{y}}_v^{init.}$.
\item \textbf{Single-step VM estimation.} The reflow-trained PeriodWave-VME generates $\hat{\mathbf{y}}_v=f_\theta(\mathbf{y}_r\mid\hat{\mathbf{y}}_v^{init.})$ by a single Euler step over the flow interval $s\in[0,1]$ (NFE$=1$). RM padding matches the VM-channel dimension as defined above. Delay estimation and pseudo-target construction are training procedures and are not performed at inference.
\item \textbf{Spatial-statistics estimation.} We concatenate the real and generated channels as $\hat{\mathbf{y}}=[\mathbf{y}_r,\hat{\mathbf{y}}_v]$. The MC-SE model processes these augmented noisy signals to obtain the speech estimate used to construct the time-frequency masks and speech/noise covariance matrices.
\item \textbf{Beamforming.} The resulting statistics determine the MCWF or MVDR coefficients and enhanced reference-channel output $\hat{\mathbf{x}}_{ref}^{bf}$, following Appendix~\ref{app:beamformers}.
\end{enumerate}
\endgroup
The guidance cue and estimated VM signals are model outputs derived from the RM signals, rather than additional observed inputs. In the end-to-end VM-SE variant, the augmented noisy signals are processed by MC-SE to produce the enhanced speech directly, without the back-end beamformer.

\section{Experiments}
\subsection{Dataset}
We built our dataset using the Interspeech 2020 Deep Noise Suppression (DNS) Challenge corpus for both speech and noise sources \cite{reddy2020interspeech}. The corpus was randomly split into 50,000, 2,000, and 3,000 clips, each with a duration of 10 s for training, validation, and testing, respectively. For room acoustics simulation, we used \textit{Pyroomacoustics} \cite{scheibler2018pyroomacoustics} and generated room impulse responses (RIRs) using the image-source method with an image-source order of six. The room dimensions (length, width, height) were uniformly sampled from $[3,10]$, $[3,10]$, and $[2,5]$ m, respectively, while the absorption coefficient was drawn uniformly from $[0.1,0.5]$. The resulting reverberation times (RT60), determined by the room dimensions and sampled absorption coefficients, ranged from $0.15$ to $1.75$ s. SNR and signal-to-interference ratio (SIR) were set in $[-10,5]$ dB. All sources were placed at a distance of $[0.5,2.5]$ m from the array center. We employed a 16-channel spherical microphone array with a radius of 0.1 m. Among the 16 microphones, we designated four channels as the RMs, forming a tetrahedral sub-array (indices $1,5,9,13$), while the remaining 12 channels served as VMs. We considered field-of-view (FoV) speech enhancement \cite{xufovnet}. In the FoV setting, the target source is constrained to the frontal direction of the array, with the FoV defined as $\pm 20^\circ$ around the front axis in both azimuth and elevation. Up to five additional interfering talkers were randomly positioned outside the FoV region within $[0.5,2.5]$\,m from the array center. The number of interfering talkers and the level of babble noise were uniformly sampled from $[0,5]$.

\subsection{Experimental setting}
We used a sampling rate of 16\,kHz with an STFT window size and hop size of 16\,ms and 8\,ms, respectively. As the MC-SE backbone, we adopted SpatialNet-small \cite{quan2024spatialnet}, which provides strong performance in an end-to-end formulation, and employed the multi-channel Wiener filter (MCWF) \cite{benesty2008microphone} as the mask-based beamformer. The output of the MC-SE model is used as a proxy for the MCWF source mask. To aggregate the temporal statistics of spatial covariance matrix estimation, we used a windowed mean computed with a rectangular window over 25 frames. All models were trained with SARL-S. In Virtual-Flow, the lightweight Spatial-Magnifier used three blocks with channel dimensions of $[64,48,32]$. For PeriodWave-VME, we used periods of $[1,2,4,8,16,32]$ and an embedding dimension of 256. We first pre-trained the lightweight Spatial-Magnifier without PeriodWave-VME, and then trained PeriodWave-VME from scratch while leveraging the pre-trained Spatial-Magnifier. Each stage was trained for 100 epochs using the Adam optimizer with a learning rate of $5\times10^{-4}$. Training was performed on 32 NVIDIA H100 GPUs with a global batch size of 64. The weighting coefficients for the neural-VME and BF losses are set to a ratio of 0.3:0.7, respectively. 

Evaluation metrics for neural-VME include scale-invariant signal-to-distortion ratio (SI-SDR, dB) \cite{le2019sdr} and SNR (dB) as time-domain metrics, multi-resolution STFT (M-STFT) \cite{yamamoto2020parallel} and log spectral distance (LSD) \cite{gray2003distance} as STFT-domain metrics since neural-VME is a task of synthesizing noisy speech. The performance of neural-VME was evaluated exclusively in a supervised manner. Then, we evaluated the speech enhancement performance using SI-SDR, SNR, perceptual evaluation of speech quality (PESQ) \cite{rix2001perceptual}, and short-time objective intelligibility (STOI, \%) \cite{taal2010short}. The model size and computational complexity were compared using parameter size and multiply-accumulate operations per second (MAC/s), respectively. For the model comparison, Spatial-Magnifier and semi-supervised Virtual-Flow were each evaluated over three independent runs with different random seeds; their VM-BF results are reported as mean $\pm$ standard deviation in the corresponding table entries.

\subsection{Results}
\subsubsection{Ablation study}
\begin{table*}[!b]
\vspace{-1em}
    \centering
    \begin{minipage}{0.48\textwidth}
        \centering
        \caption{Effect of pre-training neural-VME}
        \label{tab:vme_init}
        \vspace{-0.5em}
        \setlength{\tabcolsep}{1pt}
        \renewcommand{\arraystretch}{1}
        \scalebox{0.78}{
        \begin{tabular}{cccccc}
        \toprule
        Training & pre-training &
        \multicolumn{4}{c}{VM-BF} \\ \cmidrule(lr){3-6}
        method & neural-VME & SI-SDR$\uparrow$ & SNR$\uparrow$ & PESQ$\uparrow$ & STOI$\uparrow$ \\
        \midrule
        \multirow{2}{*}{Supervised}
        & \ding{56}
        & \textbf{10.2} & \textbf{10.8} & 2.72 & \textbf{90.1} \\
        & \ding{52}
        & 10.1 & 10.7 & \textbf{2.74} & 90.0 \\
        \midrule
        \multirow{2}{*}{Semi-supervised}
        & \ding{56}
        & 9.6 & 10.3 & 2.71 & 89.3 \\
        & 
        \cellcolor{gray!12}\ding{52}
        & 
        \cellcolor{gray!12}\textbf{10.3} & 
        \cellcolor{gray!12}\textbf{10.8} & 
        \cellcolor{gray!12}\textbf{2.79} & 
        \cellcolor{gray!12}\textbf{90.2} \\
        \bottomrule
        \end{tabular}}
    \end{minipage}
    \hfill
    \begin{minipage}{0.48\textwidth}
        \centering
        \caption{Semi-supervised learning methods}
        \label{tab:unsupervised}
        \vspace{-0.5em}
        \setlength{\tabcolsep}{1pt}
        \renewcommand{\arraystretch}{1}
        \scalebox{0.78}{
        \begin{tabular}{lcccc}
        \toprule
        \multirow{2}{*}{Method} & \multicolumn{4}{c}{VM-BF} \\
        \cmidrule(lr){2-5}
         & SI-SDR$\uparrow$ & SNR$\uparrow$ & PESQ$\uparrow$ & STOI$\uparrow$ \\
        \midrule
        w/o semi-supervised loss & 9.2 & 9.9 & 2.70 & 88.8 \\ \hline
        learnable weight & 6.5 & 7.6 & 2.34 & 82.5 \\
        PCC loss \cite{wang2024unsupervised} & 9.9 & 10.6 & 2.77 & 90.1 \\
        $\beta$ divergence \cite{yamaoka2019cnn} & 9.9 & 10.5 & 2.75 & 89.9 \\
        \rowcolor{gray!12}
        \textbf{Fixed weight} & \textbf{10.3} & \textbf{10.8} & \textbf{2.79} & \textbf{90.2} \\ \hline
        \end{tabular}}
    \end{minipage}
\end{table*}

\textbf{Pre-training neural-VME}
We examined the effect of pre-training the neural-VME model by comparing two training strategies: using a pre-trained lightweight Spatial-Magnifier as a guidance cue for PeriodWave-VME, versus training both models jointly from scratch. In the supervised setting, the lightweight Spatial-Magnifier was trained using a multi-task loss \cite{segawa2024neural}, whereas in the semi-supervised setting, only the BF loss was used. As shown in Table~\ref{tab:vme_init}, pre-training improves all reported metrics in the semi-supervised setting, whereas in the supervised setting PESQ improves slightly while SI-SDR, SNR, and STOI decrease slightly. The larger gains under semi-supervised learning suggest that pre-trained guidance helps compensate for the absence of ground-truth VM targets. The guidance and beamforming-loss ablations are reported in Appendix~\ref{app:guidance_bf_ablation}. The other ablation studies, including changing the beamformer, a lightweight model, reflow training, and changing the array geometry, are described in Appendices~\ref{app:exp_bf}, \ref{app:lightweight}, \ref{app:reflow}, and \ref{app:array_geometry}, respectively.

\textbf{Semi-supervised learning}
Table \ref{tab:unsupervised} reports the ablation results for semi-supervised learning. We conducted a comparative analysis to evaluate our design choices against several alternatives: (1) omitting the semi-supervised learning loss, (2) replacing the fixed-weight pseudo target with a learnable extraction module, (3) substituting the flow matching loss with a Pearson correlation coefficient (PCC) loss \cite{wang2024unsupervised}, and (4) employing a $\beta$ divergence \cite{yamaoka2019cnn}-based pseudo target. While a certain level of reconstruction is still achievable even without using the semi-supervised learning loss, the performance is substantially worse than that of the pseudo target-based approaches. When making the pseudo-target weights learnable, we extracted them using a \texttt{Linear--SiLU--Linear--Tanh} module. However, this strategy underperforms the variant with fixed weights. In contrast, both the $\beta$ divergence-based objective and the PCC loss achieve strong performance, demonstrating that semi-supervised flow matching training with pseudo targets is effective. However, both methods suffer from relatively low performance because they can only estimate VM signals located in-line between the RMs \cite{yamaoka2019cnn, wang2024unsupervised}. An example plot of the synthesized VM positions is shown in the Appendix \ref{app:vm_location}.

\textbf{Why does VM estimation help for beamforming?} We examine whether the improvement comes from enhanced spatial filtering or from implicit speech denoising in the generated channels. Neural-VME transforms relationships among the observed RMs into a higher-dimensional representation for covariance-based beamforming, without adding independent sensor observations. Comparisons with noisy VM references, MVDR evaluation without a Wiener post-filter, and the co-directional experiment jointly support that generated VM signals improve enhancement by strengthening spatial filtering (Appendix~\ref{app:why_vm_helps}).

\begin{table*}[!t]
\caption{Performance comparison of Virtual-Flow against various neural-VME methods.}
\label{tab:comparion_sota}
\vspace{-0.5em}
\centering
\setlength{\tabcolsep}{1.0pt}
\renewcommand{\arraystretch}{1}
\scalebox{0.685}{
\begin{tabular}{llccccccccccc}
\toprule
\multirow{2}{*}{Method} & \multirow{2}{*}{Model} & \multirow{2}{*}{VM targets} &
\multicolumn{4}{c}{Neural-VME} &
\multicolumn{4}{c}{VM-BF} &
\multicolumn{2}{c}{Complexities} \\
\cmidrule(lr){4-7}\cmidrule(lr){8-11}\cmidrule(lr){12-13}
& & &
M-STFT$\downarrow$ & LSD$\downarrow$ & SI-SDR$\uparrow$ & SNR$\uparrow$ &
SI-SDR$\uparrow$ & SNR$\uparrow$ & PESQ$\uparrow$ & STOI$\uparrow$ &
Param. & MAC/s \\
\midrule
\multicolumn{2}{l}{Unprocessed}
& -- & -- & -- & -- & --
& -11.4 & -10.3 & 1.28 & 49.6
& -- & -- \\ \hline
\multicolumn{2}{l}{SpatialNet-MCWF (4ch)}
& -- & -- & -- & -- & --
& 5.7 & 7.0 & 2.26 & 80.4
& 1.6\,M & 23.2\,G \\
\midrule
\multirow{3}{*}{Discriminative}
& Segawa \textit{et al.} \cite{segawa2024neural} & \ding{52}
& 1.78 & 3.23 & 4.4 & 5.8
& 6.7 & 8.0 & 2.38 & 83.7
& +13.1\,M & +10.3\,G \\
& Qiao \textit{et al.} \cite{qiao2025neural} & \ding{52}
& 1.57 & 2.50 & 5.1 & 6.4
& 6.0 & 7.5 & 2.19 & 80.6
& +8.2\,M & +66.0\,G \\
& Pan \textit{et al.} \cite{pan2025enhancing} & \ding{52}
& 1.42 & 2.27 & 5.2 & 6.4
& 6.4 & 7.4 & 2.35 & 83.9
& +1.6\,M & +23.2\,G \\
\midrule
\multirow{4}{*}{Generative}
& SGMSE+ \cite{richter2023speech} & \ding{52}
& 1.41 & 2.46 & 3.6 & 5.2
& 6.7 & 7.9 & 2.38 & 84.5
& +65.7\,M & +4.8\,T \\
& FlowSE \cite{lee2025flowse} & \ding{52}
& 1.41 & 2.48 & 5.5 & 6.7
& 6.7 & 8.0 & 2.40 & 84.1
& +65.7\,M & +0.8\,T \\
& Spatial-Magnifier \cite{lee2026spatial} & \ding{52}
& 1.48 & 2.83 & \textbf{6.2} & \textbf{7.3}
& 9.8$\pm$0.2 & 10.4$\pm$0.2 & 2.61$\pm$0.06 & 89.0$\pm$0.3
& +1.5\,M & +10.1\,G \\
\rowcolor{gray!12}\multicolumn{1}{c}{\cellcolor{white}}
& \textbf{Virtual-Flow} & \ding{52}
& \textbf{1.29} & \textbf{2.10} & -1.3 & 1.7
& \textbf{10.1} & \textbf{10.7} & \textbf{2.74} & \textbf{90.0}
& \textbf{+13.1\,M} & \textbf{+7.8\,G} \\
\midrule\multirow{3}{*}{Semi-supervised}
& ArrayDPS \cite{xu2025arraydps} & \ding{56}
& - & - & - & - & 5.1 & 6.5 & 2.27 & 77.3
& +405.1\,M & +2.5\,T \\
& VM-UNSSOR \cite{he2026vm} & \ding{56}
& - & - & - & - & 3.9 & 5.8 & 2.09 & 76.3
& +1.2\,M & +16.8\,G \\
\rowcolor{gray!12}\multicolumn{1}{c}{\cellcolor{white}}
& \textbf{Virtual-Flow} & \ding{56}
& - & - & - & - & \textbf{10.3$\pm$0.2} & \textbf{10.8$\pm$0.1} & \textbf{2.79$\pm$0.02} & \textbf{90.2$\pm$0.1}
& \textbf{+13.1\,M} & \textbf{+7.8\,G} \\
\midrule
\multicolumn{2}{l}{SpatialNet-MCWF (16ch)}
& -- & -- & -- & -- & --
& 12.1 & 12.4 & 2.82 & 92.8
& 1.6\,M & 23.2\,G \\
\bottomrule
\end{tabular}}
\end{table*}

\begin{table*}[!t]
\caption{Detailed neural-VME analysis in the STFT domain for Spatial-Magnifier and Virtual-Flow}
\label{tab:vme_qualtiy_analysis}
\vspace{-0.5em}
\centering
\setlength{\tabcolsep}{1pt}
\renewcommand{\arraystretch}{1.1}
\scalebox{0.78}{
\begin{tabular}{l *{12}{>{\centering\arraybackslash}p{28pt}}} 
\toprule
Model &
\multicolumn{4}{c}{Complex STFT-SNR (dB)$\uparrow$} &
\multicolumn{4}{c}{Magnitude STFT-SNR (dB)$\uparrow$} &
\multicolumn{4}{c}{IPD (rad)$\downarrow$} \\ \cmidrule(lr){1-1}\cmidrule(lr){2-5}\cmidrule(lr){6-9}\cmidrule(lr){10-13}
Frequency range (kHz) & 0--0.5 & 0.5--2 & 2--4 & 4--8
& 0--0.5 & 0.5--2 & 2--4 & 4--8
& 0--0.5 & 0.5--2 & 2--4 & 4--8 \\
\midrule
Spatial-Magnifier \cite{lee2026spatial}
& 4.9 & \textbf{4.7} & \textbf{1.2} & \textbf{0.3}
& 7.5 & \textbf{7.2} & 2.8 & 1.3
& 1.11 & \textbf{1.34} & \textbf{1.46} & \textbf{1.55} \\\rowcolor{gray!12}
\textbf{Virtual-Flow}
& \textbf{7.9} & -0.7 & -3.6 & -4.0
& \textbf{11.8} & 5.6 & \textbf{3.3} & \textbf{2.1}
& \textbf{0.47} & 1.42 & 1.59 & 1.57 \\
\bottomrule
\end{tabular}}
\vspace{-1em}
\end{table*}

\subsubsection{Comparison with existing methods}
Table~\ref{tab:comparion_sota} presents a comparison with existing discriminative \cite{segawa2024neural, qiao2025neural, pan2025enhancing}, generative \cite{richter2023speech, lee2025flowse, lee2026spatial}, and semi-supervised baselines without ground-truth VM supervision \cite{xu2025arraydps, he2026vm}. We additionally adapted generative speech enhancement models (SGMSE+, FlowSE) for the neural-VME task as baselines. For comparison with the supervised neural-VME baselines, we also trained Virtual-Flow in a supervised setting to ensure a fair comparison. The proposed Virtual-Flow achieves the strongest VM-BF performance among all methods while maintaining the lowest computational complexity among generative approaches. Notably, although Spatial-Magnifier achieves higher neural-VME performance on time-domain metrics, this does not translate into superior VM-BF performance; we analyze this discrepancy in detail below. Visual comparisons of neural-VME spectrograms are provided in Appendix~\ref{app:VME_spec}.

We compare semi-supervised Virtual-Flow with ArrayDPS \cite{xu2025arraydps} and VM-UNSSOR \cite{he2026vm}, which also omit ground-truth VM supervision while retaining clean target speech for enhancement supervision in the evaluated systems. Virtual-Flow achieves higher VM-BF performance under this shared supervision category. These results support the utility of pseudo target-based flow matching without requiring observed VM signals. Furthermore, we compared Virtual-Flow with multi-frame MCWF and an Acoustics-Informed Neural Network (AINN) in Appendix \ref{app:additional_comparisons}.

\textbf{Correlation between neural-VME and VM-BF performance} A notable observation from Table~\ref{tab:comparion_sota} is that neural-VME metrics do not always correlate with VM-BF performance. As shown in Table~\ref{tab:vme_qualtiy_analysis}, Virtual-Flow achieves stronger magnitude recovery across most frequency bands, while phase estimation remains challenging, particularly at high frequencies. Since time-domain metrics such as SI-SDR are sensitive to phase errors, they penalize Virtual-Flow disproportionately despite its superior magnitude reconstruction. Meanwhile, the improved high-frequency magnitude estimation preserves the spatial cues that frequency-domain beamforming relies on, explaining the stronger VM-BF performance. This analysis is restricted to supervised Virtual-Flow, for which neural-VME reconstruction metrics are reported. Our analysis shows that flow matching-based neural-VME achieves strong
magnitude estimation that provides sufficient spatial cues for
effective beamforming, even with limited phase accuracy.
Improving phase estimation remains a promising direction for
further gains. Spectrogram comparisons are provided in Appendix~\ref{app:spectrogram}. The generated VMs are deterministic functions of the RMs; their benefit is interpreted as learned spatial representation and covariance augmentation. Appendix~\ref{app:flow_matching} discusses this information-preserving interpretation and the rationale for single-step reflow inference.

\subsubsection{Integration of neural-VME with end-to-end model for VM-SE}
Finally, we evaluate Virtual-Flow in the VM-SE setting, where neural-VME is used to improve end-to-end speech enhancement without a back-end beamformer. This experiment uses two RM input channels, rather than the four-RM configuration used for VM-BF; the small and large SpatialNet baselines and the VM-augmented systems share this two-RM input setting. As shown in Table~\ref{tab:e2e_model}, augmenting SpatialNet-small (2ch) with the supervised Spatial-Magnifier improves speech enhancement performance. Notably, the semi-supervised Virtual-Flow achieves comparable VM-SE performance despite being trained without VM observations, and even outperforms the supervised Spatial-Magnifier in SI-SDR (-0.7 vs -1.0~dB). Furthermore, SpatialNet-small with Virtual-Flow approaches the performance of SpatialNet-large (2ch) at substantially lower computational cost (31.0 vs 119.0~G MAC/s), demonstrating that virtual spatial cues can serve as an efficient alternative to scaling model capacity. The significance of this experiment lies in demonstrating the feasibility of semi-supervised VM-SE, achieving performance gains with marginal computational overhead. Appendix~\ref{app:realman} provides real-recorded and RIR-modeling experiments, and Appendix~\ref{app:sensitivity} documents the limitations under delay errors, strong interference, reverberation, and speaker motion. Also, experiments for both the real-time and array-agnostic models are provided in Appendix~\ref{app:rt_aga}.
\begin{table*}[!t]
\caption{Comparison VM-SE results with end-to-end models}
\label{tab:e2e_model}
\vspace{-0.5em}
\centering
\setlength{\tabcolsep}{1pt}
\renewcommand{\arraystretch}{1.1}
\scalebox{0.78}{
\begin{tabular}{clccccccc}
\toprule
\multirow{2}{*}{Scenario} & \multirow{2}{*}{Method} & \multirow{2}{*}{VM targets} & \multicolumn{4}{c}{VM-SE} & \multicolumn{2}{c}{Efficiency} \\
\cmidrule(lr){4-7} \cmidrule(lr){8-9}
& & & SI-SDR$\uparrow$ & SNR$\uparrow$ & PESQ$\uparrow$ & STOI$\uparrow$ & Param. & MAC/s \\
\midrule
& SpatialNet-small (2ch) & - & -3.0 & 4.1 & 1.90 & 66.9 & 1.6\,M & 23.2\,G \\ \cline{2-9}
& \quad + Spatial-Magnifier & \ding{52} & -1.0 & 5.6 & 2.15 & 71.1 & \textbf{+1.5\,M} & +10.1\,G \\
\rowcolor{gray!12} \cellcolor{white} \multirow{-1}{*}{End-to-end model} 
& \quad + Virtual-Flow & \ding{56} & -0.7 & 5.3 & 2.09 & 70.6 & +13.1\,M & \textbf{+7.8\,G} \\
& SpatialNet-large (2ch) & - & \textbf{0.2} & \textbf{5.7} & \textbf{2.16} & \textbf{72.7} & 7.9\,M & 119.0\,G \\ \cline{2-9}
& SpatialNet-small (16ch) & - & 12.8 & 13.0 & 2.91 & 93.7 & 1.6\,M & 23.2\,G \\
\bottomrule
\end{tabular}}
\vspace{-1.5em}
\end{table*}

\section{Conclusion}
In this work, we presented Virtual-Flow, a semi-supervised flow matching-based framework for neural-VME. By leveraging a pseudo target-based training strategy, our approach generates effective virtual spatial information without requiring ground-truth VM signals. Extensive evaluations demonstrate that Virtual-Flow outperforms conventional neural-VME models in VM-BF and achieves competitive VM-SE performance. Furthermore, our method shows robustness across diverse beamforming back-ends and microphone array geometries. Our analysis shows that strong magnitude estimation provides sufficient spatial cues for effective beamforming, even with limited phase accuracy.

\textbf{Broader Impacts} The proposed framework has significant potential to democratize high-quality spatial audio processing for consumer electronics. By enabling sophisticated multichannel speech enhancement on devices with limited physical constraints, such as AR glasses and hearing aids, Virtual-Flow can improve the accessibility of assistive listening technologies and enhance the user experience in immersive communication. Furthermore, the semi-supervised nature of our method reduces the dependency on large-scale, synchronized multichannel datasets, thereby lowering the environmental and computational costs associated with data collection and supervised pre-training.

\textbf{Limitations and Future Work} Despite its effectiveness, our study has certain limitations. While the proposed reflow training enables single-step inference, the overall computational complexity, though lower than most generative models, still poses a challenge for ultra-low-power embedded processing. Additionally, the current semi-supervised pseudo target generation assumes a relatively simplified acoustic relationship between sources and RM signals, which may require further refinement for extremely reverberant or dynamic environments. Finally, while Virtual-Flow achieves strong magnitude recovery, phase estimation remains limited, particularly at high frequencies, leaving room for further beamforming improvements. Future research will focus on improving phase estimation and optimizing the model architecture for hardware-level acceleration.

\newpage

\bibliographystyle{unsrt}
\bibliography{mybib}


\appendix
\onecolumn
\section*{Appendix / Technical Appendices and Supplementary Material}
This appendix is organized as follows:
\begin{itemize}
    \item Appendix \ref{app:beamformers} provides detailed description of beamforming for VM-BF.
    \item Appendix \ref{app:flow_matching} explains detail description about relations between flow matching and neural-VME.
    \item Appendix \ref{app:why_vm_helps} analyses why generated VM signals improve enhancement and examines the implicit post-filtering hypothesis.
    \item Appendix \ref{app:loss} presents supervised learning loss for Virtual-Flow.
    \item Appendix \ref{app:additional_comparisons} compares Virtual-Flow with multi-frame MCWF and acoustics-informed neural network.
    \item Appendix \ref{app:guidance_bf_ablation} examines the effects of guidance and beamforming supervision.
    \item Appendix \ref{app:reflow} examines the effect of reflow training.
    \item Appendix \ref{app:exp_bf} shows applicability to different adaptive beamformers.
    \item Appendix \ref{app:array_geometry} shows applicability to different microphone array geometries.
    \item Appendix \ref{app:lightweight} provides experimental results with lightweight Spatial-Magnifier.
    \item Appendix \ref{app:realman} presents real-world experiments and examines the effect of RIR modeling.
    \item Appendix \ref{app:sensitivity} examines sensitivity to delay estimation, interference, reverberation, and speaker motion.
    \item Appendix \ref{app:rt_aga} provides a real-time and array-agnostic variant for VM-SE and experimental results.
    \item Appendix \ref{app:vm_location} provides analysis for estimated VM locations.
    \item Appendix \ref{app:VME_spec} presents estimated VM spectrograms from various neural-VME models.
    \item Appendix \ref{app:spectrogram} presents the spectrogram comparison between Spatial-Magnifier and Virtual-Flow.
\end{itemize}

\section{Method details}
\subsection{Explanation of beamforming procedure}\label{app:beamformers}
In the VM-BF framework, the neural-VME model is integrated with a frequency-domain spatial filter to produce the final enhanced speech. The beamformed output of the reference channel $\hat{X}_{ref}^{bf}(t, f) \in \mathbb{C}$ is obtained by applying the adaptive beamforming weights $\mathbf{W}(t, f) \in \mathbb{C}^{M \times 1}$ to the augmented multi-channel input $\hat{\mathbf{Y}}(t, f) \in \mathbb{C}^{M \times 1}$ at each time frame $t$ and frequency bin $f$:
\begin{equation}
    \hat{X}_{ref}^{bf}(t, f) = \mathbf{W}^H(t, f) \hat{\mathbf{Y}}(t, f),
\end{equation}
where $M = M_r + M_v$ is the total number of channels, including both real and generated virtual microphones. To derive the optimal weights $\mathbf{W}(t, f)$, we evaluate two representative beamforming criteria: the Multi-Channel Wiener Filter (MCWF) and the Minimum Variance Distortionless Response (MVDR) beamformer. The following subsections detail their respective formulations and the spatial statistics employed.

\textbf{Multi-Channel Wiener Filter (MCWF)}
The objective of the MCWF is to minimize the Mean Squared Error (MSE) between the beamformer output and the direct-path reference clean speech. In our implementation, the weights are derived using recursive time-average expectations of the spatial statistics. The spatial covariance matrix of the augmented input and the cross-covariance matrix between the input and the estimated reference signal $\hat{X}_{ref}$ are computed as follows:
\begin{align}
    \mathbb{E}[\mathbf{Y}\mathbf{Y}^H](t, f) &= \frac{1}{L} \sum_{i=t-\frac{L}{2}}^{t+\frac{L}{2}} \hat{\mathbf{Y}}(i, f)\hat{\mathbf{Y}}^H(i, f), \\
    \mathbb{E}[\mathbf{Y}\hat{X}_{ref}^H](t, f) &= \frac{1}{L} \sum_{i=t-\frac{L}{2}}^{t+\frac{L}{2}} \hat{\mathbf{Y}}(i, f)\hat{X}_{ref}^H(i, f).
\end{align}
For the spatial statistics, we applied a windowed mean of length $L$ to design a time-varying adaptive beamformer. Using these expectations, the optimal weight vector for the MCWF is formulated as:
\begin{equation}
    \mathbf{W}_{\text{MCWF}}(t, f) = \mathbb{E}[\mathbf{Y}\mathbf{Y}^H](t, f)^{-1} \mathbb{E}[\mathbf{Y}\hat{X}_{ref}^H](t, f).
\end{equation}
This formulation allows the filter to dynamically adapt to the acoustic environment by leveraging the expanded spatial information provided by the virtual microphones.

\textbf{Minimum Variance Distortionless Response (MVDR)}
The MVDR beamformer aims to minimize the residual noise variance while ensuring the target speech remains undistorted. For this beamformer, we utilize spatial covariance matrices (SCMs) estimated via time-frequency (T-F) masks. Using speech and noise time-frequency masks derived from the pre-trained MC-SE output, we estimate the target speech SCM ($\mathbf{\Phi}_{XX}$) and noise SCM ($\mathbf{\Phi}_{NN}$) from the augmented observations:
\begin{align}
    \mathbf{\Phi}_{XX}(t, f) &= \frac{\sum_{i\in\mathcal{W}_t} m_X(i,f)\hat{\mathbf{Y}}(i,f)\hat{\mathbf{Y}}^H(i,f)}{\sum_{i\in\mathcal{W}_t}m_X(i,f)}, \\
    \mathbf{\Phi}_{NN}(t, f) &= \frac{\sum_{i\in\mathcal{W}_t} m_N(i,f)\hat{\mathbf{Y}}(i,f)\hat{\mathbf{Y}}^H(i,f)}{\sum_{i\in\mathcal{W}_t}m_N(i,f)},
\end{align}
where $m_X(i,f)$ and $m_N(i,f)$ are the nonnegative speech and noise masks, $\hat{\mathbf{Y}}(i,f)\in\mathbb{C}^{M\times1}$ is the augmented STFT observation, and $\mathcal{W}_t$ contains the $L$ frames used for local covariance estimation. The masks weight the observation outer products separately for speech and noise. Based on these SCMs, the time-variant MVDR filter coefficients are computed as:
\begin{equation}
    \mathbf{W}_{\text{MVDR}}(t, f) = \frac{\mathbf{\Phi}_{NN}^{-1}(t, f) \mathbf{\Phi}_{XX}(t, f)}{\text{Tr}(\mathbf{\Phi}_{NN}^{-1}(t, f) \mathbf{\Phi}_{XX}(t, f))} \mathbf{u}_{\text{ref}},
\end{equation}
where $\text{Tr}(\cdot)$ denotes the trace operator and $\mathbf{u}_{\text{ref}} \in \mathbb{R}^{M\times1}$ is a one-hot vector representing the reference microphone index. The high-fidelity virtual channels generated by Virtual-Flow enhance the spectral-spatial resolution of these SCMs, leading to superior interference suppression.

\subsection{Design motivation for flow matching in virtual microphone estimation}\label{app:flow_matching}

Conventional neural-VME methods use discriminative regression to map RM signals to VM signals. Our conditional flow-matching formulation instead learns a vector field along a linear training interpolation between the padded RM-derived initial state and the selected endpoint $\mathbf{y}_v^\star$. The endpoint is an observed VM signal in supervised training and a pseudo-target in semi-supervised training. Linear interpolation of training samples does not by itself guarantee straight trajectories for the learned ODE. Reflow \cite{liu2023flow} is used to improve single-step inference, as explained below. In the STFT domain, the model recovers high-frequency magnitude structure that can benefit beamforming, while phase estimation remains challenging.

The guidance cue $\hat{\mathbf{y}}_v^{init.}$ is computed from the RM signals by the pre-trained lightweight Spatial-Magnifier. It supplies an initial VM estimate as conditioning for the vector field, simplifying the refinement required from the RM-derived initial state. Table~\ref{tab:vme_init} shows the benefit of pre-training, and Appendix~\ref{app:guidance_bf_ablation} reports the performance and training-stability effects of removing guidance. These experiments support the usefulness of conditioning without implying that guidance alone guarantees a straight ODE trajectory.

Finally, computing the CFM loss in the STFT domain aligns the training objective with the downstream beamformer, which operates in the frequency domain. The $\ell_1$ distance on the complex STFT representation optimizes for both magnitude and phase accuracy. In practice, the model achieves strong magnitude recovery while phase estimation remains challenging, particularly at high frequencies. Nevertheless, as shown in our experiments, the improved magnitude estimation at high frequencies preserves spatial cues critical for beamforming and proves sufficient for strong downstream VM-BF performance
(Table~\ref{tab:comparion_sota}).

\paragraph{Why reflow enables single-step inference.}
In Virtual-Flow, the initial state is $\mathbf{z}=\tilde{\mathbf{y}}_r+\sigma\boldsymbol{\epsilon}$, where $\boldsymbol{\epsilon}\sim\mathcal{N}(\mathbf{0},\mathbf{I})$. Thus, for a fixed RM observation, $p_0(\mathbf{z}\mid\mathbf{y}_r)=\mathcal{N}(\tilde{\mathbf{y}}_r,\sigma^2\mathbf{I})$: the initial state preserves the structure of the padded RM signals, with an additive Gaussian perturbation. The first training stage learns a conditional vector field along interpolations from this RM-derived state to $\mathbf{y}_v^\star$, which is an observed VM signal in supervised training or a pseudo-VM target in semi-supervised training. Both endpoints are associated with the same RM observation. In our reflow stage, we fix $s=0$ and retain the original target displacement $\mathbf{y}_v^\star-\mathbf{z}$, specifically $\tilde{\mathbf{y}}_v-\mathbf{z}$ in semi-supervised training; the endpoint is not replaced by a first-stage model output. This directly trains the initial-time vector-field prediction for single-step inference. Here, our use of ``reflow'' refers to this two-stage training scheme, rather than the endpoint-recoupling procedure of Liu et al.~\cite{liu2023flow}. For an ideal trajectory with constant velocity along the path, one Euler step reaches its endpoint exactly; finite model capacity and training do not guarantee this ideal condition or a globally linear input--output mapping. At inference, we set $\boldsymbol{\epsilon}=\mathbf{0}$ and use the deterministic single-step estimate $\hat{\mathbf{y}}_v=\tilde{\mathbf{y}}_r+u_0(\tilde{\mathbf{y}}_r\mid\hat{\mathbf{y}}_v^{init.})$ over $s\in[0,1]$. The effectiveness of this NFE$=1$ approximation is supported empirically by the reflow ablation in Appendix~\ref{app:reflow}.

\paragraph{Information-theoretic perspective on Virtual Microphone Estimation.}
A Neural-VME system operating solely on observed RM signals does not increase the absolute amount of information present in the input without additional physical sensors. However, when examining the relationships among RM signals in a low-resolution array with a limited number of microphones, the spatial resolution is intrinsically low, making it remarkably difficult for the downstream backend to directly extract the necessary spatial cues. To address this challenge, Virtual-Flow operates not by creating new information out of nothing, but through the following key mechanism: Learning for explicit high-dimensional spatial representation. Virtual-Flow trains effectively and regularizes the model to exploit latent spatial information far more efficiently for multichannel speech enhancement. Instead of pre-defining VM signals in a rigid or ad-hoc manner, Virtual-Flow provides the model with greater flexibility to autonomously learn and decide which VM signals are optimal for capturing spatial representations. By leveraging the relative spatial relationships among RM signals alongside a deep generative prior, Virtual-Flow restructures implicit spatial cues latent across RMs into an explicit high-dimensional representation that the downstream beamforming backend can readily and optimally exploit. In conclusion, our work does not create entirely new information; rather, it provides a flexible regularization framework that reformulates latent spatial relationships into an explicit high-dimensional spatial representation.

\subsection{Why does VM estimation help?}\label{app:why_vm_helps}
Virtual-Flow does not create new information. Its purpose is to improve spatial filtering by transforming spatial relationships already latent in a small physical array into an explicit, higher-dimensional representation for covariance-based beamforming. With few microphones, estimating sufficiently informative spatial covariance matrices is difficult, and the enhancement model may rely heavily on spectral filtering. Virtual-Flow instead learns virtual channels optimized for downstream spatial filtering. 

A possible alternative explanation is that Virtual-Flow generates cleaner speech signals and therefore acts as implicit post-processing. We conducted three analyses to test this hypothesis. 

(a) The generated VMs retain noisy microphone characteristics: Table~\ref{tab:comparion_sota} and Appendices~\ref{app:VME_spec}--~\ref{app:spectrogram} compare the generated VMs with reference noisy VM recordings and other spatial upsampling baselines. Virtual-Flow reproduces the energy and spectrotemporal structure of the noisy reference VMs rather than producing clean target speech.

(b) Virtual-Flow improves a beamformer without a post-filter: We evaluated Virtual-Flow with a Souden-MVDR backend (Appendix~\ref{app:exp_bf}, Table~\ref{tab:mvdr}). Unlike MCWF, MVDR contains no single-channel Wiener post-filter. With identical mask and covariance-estimation procedures, Virtual-Flow improves SpatialNet-MVDR from 5.6 to 9.0 dB SI-SDR, approaching the 16-channel physical-array result of 10.0 dB. Thus, the benefit remains when post-filtering is absent. 

(c) The Virtual-Flow advantage contracts when directional separation is removed: We additionally evaluated MCWF when the target and interferer have the same direction (co-directional noise):

\begin{table}[H]
\vspace{-1em}
\centering
\caption{Effect of target--interferer direction. SI-SDR and SNR are in dB, and STOI is in \%. The separated-condition Virtual-Flow values are the reported means from the baseline comparison; co-directional uncertainty estimates are not available. NN MAC/s excludes beamforming computation.}
\label{tab:codirectional}
\setlength{\tabcolsep}{4pt}
\renewcommand{\arraystretch}{1.1}
\resizebox{\linewidth}{!}{
\scalebox{0.78}{
\begin{tabular}{lcccccccccc}
\toprule
\multirow{2}{*}{Method} & \multicolumn{4}{c}{Spatially separated} & \multicolumn{4}{c}{Co-directional} & \multicolumn{2}{c}{Efficiency} \\
\cmidrule(lr){2-5}\cmidrule(lr){6-9}\cmidrule(lr){10-11}
& SI-SDR$\uparrow$ & SNR$\uparrow$ & PESQ$\uparrow$ & STOI$\uparrow$ & SI-SDR$\uparrow$ & SNR$\uparrow$ & PESQ$\uparrow$ & STOI$\uparrow$ & Param. & NN MAC/s \\
\midrule
Unprocessed & -11.4 & -10.3 & 1.28 & 49.6 & -10.6 & -10.4 & 1.24 & 48.2 & -- & -- \\ \hline
SpatialNet-MCWF (4 RM) & 5.7 & 7.0 & 2.26 & 80.4 & 4.1 & 5.0 & 2.12 & 76.2 & 1.6\,M & 23.2\,G \\ \hline
Multi-frame MCWF (4 RM $\times$ 4 frames) & 7.2 & 7.7 & 2.48 & 87.0 & 4.8 & 5.6 & 2.18 & 78.1 & \textbf{1.6\,M} & \textbf{23.2\,G} \\
\rowcolor{gray!12}
\textbf{Virtual-Flow (4 RM + 12 VM)} & \textbf{10.3} & \textbf{10.8} & \textbf{2.79} & \textbf{90.2} & \textbf{5.3} & \textbf{6.2} & \textbf{2.21} & \textbf{79.4} & 14.7\,M & 31.0\,G \\ \hline
SpatialNet-MCWF (16 RM) & 12.1 & 12.4 & 2.82 & 92.8 & 6.8 & 7.7 & 2.42 & 84.2 & 1.6\,M & 23.2\,G \\
\bottomrule
\end{tabular}}}
\vspace{-1em}
\end{table}

When target and noise arrive from different directions, Virtual-Flow gains 4.6 dB over the 4-channel SpatialNet baseline and substantially closes the gap to the physical 16-channel array. When they are co-directional, and direct-path directional separability is greatly reduced, the gain falls to 1.2 dB. The physical 16-channel advantage contracts similarly. These experiments provide converging evidence that Virtual-Flow primarily improves spatial filtering: its benefit persists without a post-filter and diminishes when directional separation is removed. Strong post-processing addresses the objective of suppressing residual distortion after beamforming, whereas Virtual-Flow improves the spatial filtering stage itself. They address distinct stages and can therefore be combined: post-processing may still be applied after Virtual-Flow.

\section{Loss function for supervised learning}\label{app:loss}
For supervised learning, we consider two training variants. The first variant uses only the flow matching loss as the VM loss, whereas the second additionally incorporates a discriminative VM loss. In both variants, we further apply an SNR loss between the clean signal and the enhanced signal to optimize either VM-BF or VM-SE.

\textbf{Version 1 (FM-only VM loss)}
The first variant is formulated as
\begin{small}
\begin{align}
\begin{aligned}
\mathcal{L}_{\mathrm{total}}
= w_{vme}\log_{10}\mathcal{L}_{\mathrm{CFM}}(\hat{\mathbf{y}}_v,\mathbf{y}_v)
+ w_{bf}\mathcal{L}_{\mathrm{SNR}}(\hat{\mathbf{x}}_{ref},\mathbf{x}_{ref}),
\end{aligned}
\end{align}
\end{small}
where the weight ratio is set to $w_{vme}:w_{bf}=0.3:0.7$.

\textbf{Version 2 (FM + discriminative VM loss)}
The proposed framework generates VM signals via single-step flow matching even during training. Therefore, in supervised learning, we can add an SNR-based VM loss. In this variant, we use a skip connection from the pre-trained lightweight Spatial-Magnifier and define the final estimated VM signal as the sum of the Spatial-Magnifier output and the Virtual-Flow output. On this estimated VM signal, we apply an SNR loss and a multi-resolution (M)-STFT loss, while the flow matching loss is applied to the Virtual-Flow output as in the first variant. The total loss is
\begin{small}
\begin{align}
\begin{aligned}
\mathcal{L}_{\mathrm{total}}
&= w_{vme}\log_{10}\mathcal{L}_{\mathrm{CFM}}(\hat{\mathbf{y}}_v,\mathbf{y}_v)
+ w_{vme}\mathcal{L}_{\mathrm{SNR}}(\hat{\mathbf{y}}_v,\mathbf{y}_v) \\
&\quad + w_{vme}\log_{10}\mathcal{L}_{\mathrm{M\text{-}STFT}}(\hat{\mathbf{y}}_v,\mathbf{y}_v)
+ w_{bf}\mathcal{L}_{\mathrm{SNR}}(\hat{\mathbf{x}}_{ref},\mathbf{x}_{ref}),
\end{aligned}
\end{align}
\end{small}
with $w_{vme}:w_{bf}=0.1:0.7$.

\textbf{Experimental results} The comparative analysis in Table \ref{tab:mcwf} reveals that while Virtual-Flow (Version 2) achieves the highest fidelity on individual neural-VME metrics, such as a 4.3 dB SI-SDR, this does not translate proportionately into superior beamforming performance. In contrast, Version 1 yields better VM-BF results, achieving an SI-SDR of 10.1 dB, suggesting that flow-matching objectives may align more effectively with downstream spatial filtering than traditional discriminative losses. Notably, the semi-supervised Virtual-Flow (semi-supervised) achieves high VM-BF performance, outperforming both supervised variants in perceptual quality (2.79 PESQ). This underscores the efficacy of our proposed pseudo target strategy in providing robust spatial cues even in the absence of ground-truth VM signals. Consequently, these results validate that our framework can successfully bridge the performance gap between sparse 4-channel setups and dense 16-channel arrays.
\begin{table*}[!h]
\caption{Performance comparison of Virtual-Flow using supervised and semi-supervised learning}
\label{tab:mcwf}
\vspace{-0.5em}
\centering
\setlength{\tabcolsep}{1pt}
\renewcommand{\arraystretch}{1.1}
\scalebox{0.78}{
\begin{tabular}{lccccccccc}
\toprule
\multirow{2}{*}{Methods} & \multirow{2}{*}{VM targets} &
\multicolumn{4}{c}{Neural-VME} &
\multicolumn{4}{c}{VM-BF} \\
\cmidrule(lr){3-6}\cmidrule(lr){7-10}
& & M-STFT$\downarrow$ & LSD$\downarrow$ & SI-SDR$\uparrow$ & SNR$\uparrow$
& SI-SDR$\uparrow$ & SNR$\uparrow$ & PESQ$\uparrow$ & STOI$\uparrow$ \\
\midrule
SpatialNet-MCWF (4ch)
& -- & -- & -- & -- & --
& 5.7 & 7.0 & 2.26 & 80.4 \\ \midrule
\quad + Virtual-Flow (version 1)
& \ding{52} & 1.29 & 2.10 & -1.3 & 1.7
& \textbf{10.1} & \textbf{10.7} & 2.74 & 90.0 \\
\quad + Virtual-Flow (version 2)
& \ding{52} & \textbf{1.14} & \textbf{2.03} & \textbf{4.3} & \textbf{5.8}
& 9.5 & 10.1 & 2.62 & 88.5 \\
\quad + Virtual-Flow (semi-supervised)
& \ding{56} & -- & -- & -- & --
& 10.3 & 10.8 & \textbf{2.79} & \textbf{90.2} \\ \midrule
SpatialNet-MCWF (16ch)
& -- & -- & -- & -- & --
& 12.1 & 12.4 & 2.82 & 92.8 \\
\bottomrule
\end{tabular}}
\vspace{-0.5em}
\end{table*}

\section{Comparison with multi-frame beamforming and Acoustics-Informed Neural Network (AINN)}\label{app:additional_comparisons}

\textbf{Comparison with multi-frame MCWF}
Table~\ref{tab:matched_covariance_comparison} compares Virtual-Flow with multi-frame MCWF \cite{wang2021multiframe} using the same SpatialNet-small backbone and beamformer configuration. For a fair comparison, we adopted SpatialNet-small (instead of the TDCN used in~\cite{wang2021multiframe}) as the speech enhancement backbone and applied MCWF with the same beamformer configuration. Crucially, both methods operate on an identical  $16\times16$ spatial covariance matrix, i.e., multi-frame stacks 4 real mics $\times$ 4 frames, while Virtual-Flow uses 4 real $+$ 12 virtual channels, so the SCM dimension is held constant across the comparison. The only difference is how the covariance is populated: multi-frame reuses the temporal context of the same spatial samples, whereas Virtual-Flow injects learned estimates at new spatial locations. Under this matched-dimension setting, the spatial-augmentation route yields larger gains, confirming that the improvement is not merely an artifact of enlarging the SCM.

We emphasize that multi-frame MCWF and Virtual-Flow are complementary rather than competing: multi-frame augments the covariance along the temporal axis, whereas Virtual-Flow utilizes a deep generative prior to augment the SCM along the spatial axis (applying the estimated beamformer coefficients to the real-microphone signals without using extra temporal context), which is why our main comparisons in the paper are against spatial VM baselines. The two are naturally combinable into a spatio-temporally augmented SCM, a promising direction we leave for future work.

\textbf{Comparison with Acoustics-Informed Neural Network (AINN)}
Table~\ref{tab:matched_covariance_comparison} shows experimental results using the Acoustically Informed Neural Network (AINN) proposed by Zhao and Ma, implemented following their original paper~\cite{zhao2024circular}. As noted in their paper, AINN simplistically models the complex acoustic field, making it difficult to handle reverberation and prone to phase distortion under realistic room environments. In contrast, our Virtual-Flow effectively reconstructs high-frequency bands through flow matching, thereby maintaining superior performance even in these challenging environments.

Their coordinate-based AINN is constrained by the homogeneous Helmholtz equation, developed under a plane-wave beamforming formulation, and validated in a hemi-anechoic chamber. Under our noisy reverberant setting, Zhao–Ma obtains -3.4 dB SI-SDR, 1.81 PESQ, and 74.4 STOI, versus 10.3 dB, 2.79, and 90.2 for Virtual-Flow. AINN augmentation also underperforms the four-RM SpatialNet-MCWF baseline without VMs (5.7 dB SI-SDR and 80.4 STOI), showing that adding its estimated channels does not improve enhancement under this evaluation. This result alone does not identify the cause of the degradation. Zhao–Ma is substantially smaller (42.8K versus 14.7M parameters). These results demonstrate that our proposed method operates robustly with high performance even in harsh, noisy, and reverberant environments, rather than being confined to controlled conditions.

\begin{table}[!htbp]
\vspace{-1em}
\caption{VM-BF comparison with multi-frame MCWF and AINN. NN MAC/s excludes beamforming computation.}
\label{tab:matched_covariance_comparison}
\centering
\setlength{\tabcolsep}{5pt}
\renewcommand{\arraystretch}{1.1}
\scalebox{0.78}{
\begin{tabular}{lcccccc}
\toprule
Method & SI-SDR$\uparrow$ & SNR$\uparrow$ & PESQ$\uparrow$ & STOI$\uparrow$ & Param. & NN MAC/s \\
\midrule
SpatialNet-MCWF (4 RM) & 5.7 & 7.0 & 2.26 & 80.4 & 1.6\,M & 23.2\,G \\ \hline
Multi-frame MCWF (4 RM $\times$ 4 frames) \cite{wang2021multiframe} & 7.2 & 7.7 & 2.48 & 87.0 & 1.6\,M & 23.2\,G \\
Zhao--Ma AINN \cite{zhao2024circular} & -3.4 & 2.8 & 1.81 & 74.4 & \textbf{42.8\,K} & \textbf{12.0\,G} \\
\rowcolor{gray!12}
\textbf{Virtual-Flow (4 RM + 12 VM)} & \textbf{10.3} & \textbf{10.8} & \textbf{2.79} & \textbf{90.2} & 14.7\,M & 31.0\,G \\ \hline
SpatialNet-MCWF (16 RM) & 12.1 & 12.4 & 2.82 & 92.8 & 1.6\,M & 23.2\,G \\
\bottomrule
\end{tabular}}
\vspace{-0.5em}
\end{table}

\section{Additional experiments and ablations}
\subsection{Effect of guidance and beamforming loss}\label{app:guidance_bf_ablation}
We ablate the Spatial-Magnifier guidance cue and downstream beamforming loss individually in semi-supervised Virtual-Flow. Table~\ref{tab:guidance_bf_ablation} reports the resulting VM-BF performance under the baseline MCWF configuration. As a result, removing guidance reduces SI-SDR from 10.3 to 9.5\,dB and PESQ from 2.79 to 2.71. We also observed larger fluctuations during training without guidance, supporting its role in stabilizing the flow-matching model when ground-truth VM signals are unavailable. Also, removing the beamforming loss reduces SI-SDR to 8.2\,dB and PESQ to 2.63. This result supports the role of downstream supervision in aligning generated VM signals with enhancement performance, rather than optimizing pseudo-target reconstruction alone.

\begin{table}[!htbp]
\vspace{-1em}
\centering
\caption{Ablation of guidance and beamforming supervision in semi-supervised Virtual-Flow}
\label{tab:guidance_bf_ablation}
\setlength{\tabcolsep}{6pt}
\scalebox{0.78}{
\begin{tabular}{lcccc}
\toprule
Method & SI-SDR$\uparrow$ & SNR$\uparrow$ & PESQ$\uparrow$ & STOI$\uparrow$ \\
\midrule
w/o Guidance & 9.5 & 10.1 & 2.71 & 89.3 \\
w/o Beamforming loss & 8.2 & 8.7 & 2.63 & 88.2 \\
\rowcolor{gray!12}
\textbf{Virtual-Flow} & \textbf{10.3} & \textbf{10.8} & \textbf{2.79} & \textbf{90.2} \\
\bottomrule
\end{tabular}}
\vspace{-0.5em}
\end{table}

\subsection{Effect of reflow training}\label{app:reflow}
To isolate the contribution of reflow training, we trained the same Virtual-Flow architecture in a supervised setting under three training schemes: a discriminative variant, a flow matching variant, and flow matching with reflow. The discriminative variant removes the time-step embedding and replaces the flow matching objective with an SNR loss on the VM signal. The flow matching variant uses time-step embedding and is trained only with the flow matching loss, and follows PeriodWave-style inference with $\mathrm{NFE}=16$. Reflow training performs Virtual-Flow training in two stages and enables single-step inference with $\mathrm{NFE}=1$. Empirically, the reflow-trained model achieves the highest VM-BF performance. In contrast, the other two variants yield worse Neural-VME quality, which also leads to substantially lower VM-BF performance. Consequently, since the proposed model was designed for reflow training, it suggests that parameter adjustment or model redesign is necessary for conventional flow matching.

\begin{table*}[!h]
\caption{Effect of reflow training}
\vspace{-0.5em}
\centering
\setlength{\tabcolsep}{1pt}
\renewcommand{\arraystretch}{1}
\scalebox{0.78}{
\begin{tabular}{lcccccccccc}
\toprule
\multirow{2}{*}{Method} &
\multicolumn{4}{c}{Neural-VME} &
\multicolumn{4}{c}{VM-BF} &
\multicolumn{2}{c}{Training / inference} \\
\cmidrule(lr){2-5}\cmidrule(lr){6-9}\cmidrule(lr){10-11}
& M-STFT$\downarrow$ & LSD$\downarrow$ & SI-SDR$\uparrow$ & SNR$\uparrow$
& SI-SDR$\uparrow$ & SNR$\uparrow$ & PESQ$\uparrow$ & STOI$\uparrow$
& Stages & NFE \\
\midrule
Discriminative & 2.47 & 2.69 & -32.1 & -3.3 & 6.7 & 7.9 & 2.35 & 84.4 & - & - \\
Flow matching & 8.40 & 5.12 & -30.5 & -12.3 & 7.6 & 8.5 & 2.44 & 83.4 & 1 & 16 \\
\rowcolor{gray!12}
\textbf{FM+reflow} & \textbf{1.29} & \textbf{2.10} & \textbf{-1.3} & \textbf{1.7} & \textbf{10.1} & \textbf{10.7} & \textbf{2.74} & \textbf{90.0} & 2 & 1 \\
\bottomrule
\end{tabular}}
\end{table*}



\subsection{Applicability to different adaptive beamformers}\label{app:exp_bf}
Tables \ref{tab:mcwf} and \ref{tab:mvdr} investigate the generalization of the proposed Virtual-Flow when integrated with SpatialNet-based beamforming back-ends under different beamformers. In Table \ref{tab:mcwf} (MCWF back-end), augmenting SpatialNet-MCWF (4ch) with Virtual-Flow (version~1) yields a substantial improvement. This result indicates that, using only four real microphones, the proposed approach can close a large portion of the gap to the 16-channel upper bound (SpatialNet-MCWF (16ch)). Moreover, the pseudo target-based semi-supervised variant attains high VM-BF performance, demonstrating that meaningful gains are achievable even without access to VM observations. Notably, version~2 can produce higher Neural-VME performance while degrading VM-BF performance, suggesting that improving neural-VME signal fidelity does not necessarily translate into better beamforming; instead, the estimated VM signals must be aligned with downstream VM-BF objectives.

\begin{table*}[!h]
\caption{Performance comparison of Virtual-Flow with MVDR back-end}
\label{tab:mvdr}
\vspace{-0.5em}
\centering
\setlength{\tabcolsep}{1pt}
\renewcommand{\arraystretch}{1.1}
\scalebox{0.78}{
\begin{tabular}{lccccccccc}
\toprule
\multirow{2}{*}{Methods} & \multirow{2}{*}{VM targets} &
\multicolumn{4}{c}{Neural-VME} &
\multicolumn{4}{c}{VM-BF} \\
\cmidrule(lr){3-6}\cmidrule(lr){7-10}
& & M-STFT$\downarrow$ & LSD$\downarrow$ & SI-SDR$\uparrow$ & SNR$\uparrow$
& SI-SDR$\uparrow$ & SNR$\uparrow$ & PESQ$\uparrow$ & STOI$\uparrow$ \\
\midrule
SpatialNet-MVDR (4ch)
& -- & -- & -- & -- & --
& 5.6 & 6.9 & 2.33 & 80.4 \\ \midrule
\quad + Virtual-Flow (version 1)
& \ding{52} & 4.10 & 3.84 & -10.8 & -7.2
& 8.7 & 9.5 & 2.66 & 87.6 \\
\quad + Virtual-Flow (version 2)
& \ding{52} & \textbf{1.14} & \textbf{1.98} & \textbf{3.4} & \textbf{5.1}
& 8.1 & 8.9 & 2.54 & 85.6 \\
\quad + Virtual-Flow (semi-supervised)
& \ding{56} & -- & -- & -- & --
& \textbf{9.0} & \textbf{9.7} & \textbf{2.67} & \textbf{87.8} \\ \midrule
SpatialNet-MVDR (16ch)
& -- & -- & -- & -- & --
& 10.0 & 10.5 & 2.73 & 89.8 \\
\bottomrule
\end{tabular}
}
\end{table*}

Table \ref{tab:mvdr} evaluates robustness to the choice of beamformer by replacing MCWF with MVDR. Similar to the MCWF case, SpatialNet-MVDR (4ch) with Virtual-Flow (version~1) improves VM-BF performance. The semi-supervised pseudo target setting also maintains strong VM-BF performance, confirming that the VM signals generated by Virtual-Flow consistently enhance spatial filtering beyond a specific beamforming formulation. These observations support that the key advantage of our approach stems from increasing effective spatial resolution via virtualization, rather than exploiting beamformer-specific properties.

\subsection{Applicability to different microphone array geometries}\label{app:array_geometry}
Furthermore, we validated the method on a simulated 7-channel array generated using measured Array Transfer Functions (ATFs) from an array comprising 5 microphones mounted in smart glasses (RM) and 2 channels representing HRTF responses (VM). Table \ref{tab:ar_glasses} examines generalization across array geometries using an AR-glasses-like configuration, where only three RM channels are available. The baseline SpatialNet-MCWF (3ch) with Virtual-Flow substantially improves recovery, with version~2 achieving the highest VM-BF performance. This result suggests that the optimal neural-VME characteristics can vary with geometry, yet the proposed framework remains effective under realistic, resource-limited wearable arrays.
\begin{table*}[!h]
\vspace{-0.5em}
\caption{Performance of Virtual-Flow with an AR-glasses-like array}
\label{tab:ar_glasses}
\vspace{-0.5em}
\centering
\setlength{\tabcolsep}{1pt}
\renewcommand{\arraystretch}{1.1}
\scalebox{0.78}{
\begin{tabular}{lccccccccc}
\toprule
\multirow{2}{*}{Methods} & \multirow{2}{*}{VM targets} &
\multicolumn{4}{c}{Neural-VME} &
\multicolumn{4}{c}{VM-BF} \\
\cmidrule(lr){3-6}\cmidrule(lr){7-10}
& & M-STFT$\downarrow$ & LSD$\downarrow$ & SI-SDR$\uparrow$ & SNR$\uparrow$
& SI-SDR$\uparrow$ & SNR$\uparrow$ & PESQ$\uparrow$ & STOI$\uparrow$ \\
\midrule
SpatialNet-MCWF (3ch)
& -- & -- & -- & -- & --
& 2.5 & 4.8 & 1.92 & 72.6 \\ \midrule
\quad + Virtual-Flow (version 1)
& \ding{52} & 1.69 & \textbf{2.17} & -12.0 & -1.8
& 5.8 & 7.1 & 2.27 & 83.2 \\
\quad + Virtual-Flow (version 2)
& \ding{52} & \textbf{1.58} & 2.31 & \textbf{-1.7} & \textbf{3.0}
& \textbf{6.0} & \textbf{7.3} & \textbf{2.32} & \textbf{82.8} \\
\quad + Virtual-Flow (semi-supervised)
& \ding{56} & -- & -- & -- & --
& 5.5 & 6.8 & 2.26 & 82.7 \\ \midrule
SpatialNet-MCWF (7ch)
& -- & -- & -- & -- & --
& 7.3 & 8.3 & 2.36 & 85.9 \\
\bottomrule
\end{tabular}
}
\vspace{-0.5em}
\end{table*}

\subsection{Lightweight model study}\label{app:lightweight}
We next compare the Spatial-Magnifier \cite{lee2026spatial} with the proposed lightweight Spatial-Magnifier, and further evaluate the effect of replacing the lightweight Spatial-Magnifier in Virtual-Flow with the Spatial-Magnifier. From the results (Table \ref{tab:lightweight}), while the lightweight Spatial-Magnifier exhibits slightly degraded performance relative to the original Spatial-Magnifier, it provides a substantial reduction in parameter count and computational cost. Interestingly, substituting the Spatial-Magnifier with the lightweight Spatial-Magnifier within Virtual-Flow improves VM-BF performance. Although the proposed Virtual-Flow has a larger parameter size than the Spatial-Magnifier baseline, its computational complexity is lower, suggesting that the proposed design achieves better performance with reduced computational complexity.

\begin{table*}[!h]
\vspace{-0.5em}
\caption{Lightweight model study}
\label{tab:lightweight}
\vspace{-0.5em}
\centering
\setlength{\tabcolsep}{1pt}
\renewcommand{\arraystretch}{1}
\scalebox{0.78}{
\begin{tabular}{cccccccccccc}
\toprule
\multicolumn{2}{c}{Method}&
\multicolumn{4}{c}{Neural-VME} &
\multicolumn{4}{c}{VM-BF} &
\multicolumn{2}{c}{Complexities} \\
\cmidrule(lr){3-6}\cmidrule(lr){7-10}\cmidrule(lr){11-12}
Spatial-Magnifier & PeriodWave-VME& M-STFT$\downarrow$ & LSD$\downarrow$ & SI-SDR$\uparrow$ & SNR$\uparrow$
& SI-SDR$\uparrow$ & SNR$\uparrow$ & PESQ$\uparrow$ & STOI$\uparrow$
& Param. & MAC/s \\
\midrule
Lee \textit{et al.} \cite{lee2026spatial} & \ding{56}
& 1.48 & 2.83 & \textbf{6.2} & \textbf{7.3}
& 9.8$\pm$0.2 & 10.4$\pm$0.2 & 2.61$\pm$0.06 & 89.0$\pm$0.3
& 1.5\,M & 10.1\,G \\
Ours & \ding{56}
& 1.58 & 3.25 & 6.1 & 7.2
& 9.8 & 10.4 & 2.65 & 89.0
& \textbf{0.25\,M} & \textbf{2.0\,G} \\
Lee \textit{et al.} \cite{lee2026spatial} & \ding{52}
& 1.30 & \textbf{2.05} & -1.4 & 1.6
& 10.0 & 10.7 & 2.70 & 89.7
& 14.3\,M & 15.9\,G \\
\rowcolor{gray!12}
Ours & \ding{52}
& \textbf{1.29} & 2.10 & -1.3 & 1.7
& \textbf{10.1} & \textbf{10.7} & \textbf{2.74} & \textbf{90.0}
& 13.1\,M & 7.8\,G \\
\bottomrule
\end{tabular}}
\end{table*}

\section{Real-world experiments and RIR modeling}\label{app:realman}
To evaluate Virtual-Flow beyond the simulated acoustic conditions used in the main experiments, we examine two complementary settings: training and evaluation on real-recorded multichannel speech, and evaluation under more detailed room impulse response (RIR) simulations.

\subsection{Real-recorded speech enhancement}
We conduct training and evaluation on RealMAN \cite{yang2024realman}, a real-recorded microphone array dataset that includes both static and moving speakers. Table~\ref{tab:realman} compares the four-channel SpatialNet-MCWF baseline, its augmentation with Virtual-Flow, and the sixteen-channel physical-array reference. Virtual-Flow is trained without ground-truth VM recordings, while the clean reference speech remains available for the downstream enhancement objective. From the 32-channel RealMAN array, we selected channels $\{1,9,17,25\}$ as RMs and designated channels $\{3,5,7,11,13,15,19,21,23,27,29,31\}$ as VMs; the 16-channel physical-array baseline uses all odd-numbered channels from 1 to 31 (one-based indexing).

\begin{table}[!htbp]
\vspace{-1.0em}
\centering
\caption{Speech enhancement results on RealMAN for static and moving speakers. SIG, BAK, and OVL denote signal quality, background quality, and overall quality scores, respectively.}
\label{tab:realman}
\setlength{\tabcolsep}{6pt}
\renewcommand{\arraystretch}{1.1}
\scalebox{0.78}{
\begin{tabular}{lcccccc}
\toprule
Method & SI-SDR$\uparrow$ & PESQ$\uparrow$ & STOI$\uparrow$ & SIG$\uparrow$ & BAK$\uparrow$ & OVL$\uparrow$ \\
\midrule
\multicolumn{7}{c}{(a) Static speakers} \\
\midrule
Unprocessed & -9.9 & 1.14 & 74.0 & 2.01 & 1.73 & 1.52 \\
SpatialNet-MCWF (4ch) & 4.3 & 1.77 & 85.2 & 2.84 & 2.38 & 2.02 \\
\rowcolor{gray!12}
\quad \textbf{+ Virtual-Flow} & \textbf{5.5} & \textbf{1.87} & \textbf{87.4} & \textbf{2.92} & \textbf{2.53} & \textbf{2.20} \\
SpatialNet-MCWF (16ch) & 6.6 & 1.95 & 88.9 & 3.08 & 2.61 & 2.30 \\
\midrule
\multicolumn{7}{c}{(b) Moving speakers} \\
\midrule
Unprocessed & -9.0 & 1.10 & 72.0 & 1.80 & 1.54 & 1.37 \\
SpatialNet-MCWF (4ch) & 1.6 & 1.42 & 81.0 & 2.59 & 2.32 & 1.96 \\
\rowcolor{gray!12}
\quad \textbf{+ Virtual-Flow} & \textbf{3.1} & \textbf{1.65} & \textbf{83.9} & \textbf{2.81} & \textbf{2.41} & \textbf{2.03} \\
SpatialNet-MCWF (16ch) & 4.1 & 1.81 & 84.8 & 2.90 & 2.52 & 2.16 \\
\bottomrule
\end{tabular}}
\vspace{-0.5em}
\end{table}

Virtual-Flow improves the four-channel baseline in both conditions. For static speakers, SI-SDR increases from 4.3 to 5.5\,dB, PESQ from 1.77 to 1.87, and STOI from 85.2 to 87.4\%. For moving speakers, the corresponding scores increase from 1.6 to 3.1\,dB, 1.42 to 1.65, and 81.0 to 83.9\%. SIG, BAK, and OVL also improve in both settings. These results demonstrate that the benefit of VM augmentation extends to real recordings and natural speaker motion when real-recorded data are included during training.

Speaker motion nevertheless presents a limitation for pseudo-target construction: superposing time-shifted recordings can produce virtual trajectories that are not physically realizable. Our beamformer estimates SCMs using a windowed mean over 25 frames, which smooths short-term temporal fluctuations. This averaging may help the downstream filter tolerate fluctuations in the generated channels, although the RealMAN results alone do not isolate its contribution. The findings support the practical applicability of Virtual-Flow under real-recorded conditions, rather than establishing that a model trained only on simulated static scenes generalizes without adaptation.

\subsection{Effect of RIR modeling}
The main simulated experiments use the image-source method (ISM) with a maximum reflection order of six. This approximation may underrepresent late reverberation, particularly for rooms with long reverberation times. To assess sensitivity to this modeling choice, we evaluate the model trained with sixth-order ISM on test data generated using sixteenth-order ISM and hybrid ray-tracing, which provide more detailed reverberation modeling. Table~\ref{tab:rir_modeling} reports these results alongside the sixth-order ISM setting.

\begin{table}[!htbp]
\vspace{-1.0em}
\centering
\caption{Effect of RIR modeling on Virtual-Flow. The model is trained using sixth-order ISM and evaluated under each simulation setting.}
\label{tab:rir_modeling}
\setlength{\tabcolsep}{8pt}
\renewcommand{\arraystretch}{1.1}
\scalebox{0.78}{
\begin{tabular}{lcccc}
\toprule
Test RIR simulation & SI-SDR$\uparrow$ & SNR$\uparrow$ & PESQ$\uparrow$ & STOI$\uparrow$ \\
\midrule
Sixth-order ISM & 10.3 & 10.8 & 2.79 & 90.2 \\
Sixteenth-order ISM & 9.4 & 10.1 & 2.71 & 89.9 \\
Hybrid ray-tracing & 8.7 & 9.4 & 2.63 & 88.1 \\
\bottomrule
\end{tabular}}
\vspace{-0.5em}
\end{table}

SI-SDR decreases from 10.3\,dB under sixth-order ISM to 9.4\,dB under sixteenth-order ISM and 8.7\,dB under hybrid ray-tracing. The other metrics also decline, indicating sensitivity to the mismatch between training and test acoustic simulations. These results motivate more realistic reverberation modeling during training. Together with the RealMAN experiments, they show both the limitations of the original simulation setup and the gains obtainable when real-recorded acoustic conditions, including moving speakers, are represented during training.

\section{Sensitivity to delay estimation and acoustic conditions}\label{app:sensitivity}
We evaluate the sensitivity of Virtual-Flow to delay-estimation errors, interference strength, reverberation time, and speaker motion. The model is trained on the simulated dataset generated using sixth-order ISM. Table~\ref{tab:sensitivity} reports VM-BF performance under each condition, together with the unprocessed input and the average Virtual-Flow performance for reference. Moving-speaker test scenes are simulated using gpuRIR.

\begin{table}[!htbp]
\vspace{-1em}
\centering
\caption{Sensitivity of Virtual-Flow to delay-estimation errors and acoustic conditions. The model is trained using sixth-order ISM. SI-SDR and SNR are reported in dB, and STOI in \%.}
\label{tab:sensitivity}
\setlength{\tabcolsep}{6pt}
\renewcommand{\arraystretch}{1.1}
\scalebox{0.78}{
\begin{tabular}{llcccc}
\toprule
Factor & Condition & SI-SDR$\uparrow$ & SNR$\uparrow$ & PESQ$\uparrow$ & STOI$\uparrow$ \\
\midrule
Unprocessed & -- & -11.4 & -10.3 & 1.28 & 49.6 \\
\rowcolor{gray!12}
Virtual-Flow & Average performance & 10.3 & 10.8 & 2.79 & 90.2 \\
\midrule
\multirow{2}{*}{Delay-estimation error}
& Low ($<0.625$\,ms) & 10.7 & 11.3 & 2.82 & 90.9 \\
& High ($\geq0.625$\,ms) & 7.6 & 8.0 & 2.54 & 87.0 \\
\midrule
\multirow{2}{*}{Interference strength}
& Weak interferers (SIR $\geq0$\,dB) & 15.2 & 15.7 & 3.01 & 95.5 \\
& Strong interferers (SIR $<0$\,dB) & 7.0 & 7.5 & 2.67 & 87.3 \\
\midrule
\multirow{3}{*}{Reverberation time}
& Low (RT60 $<0.4$\,s) & 10.6 & 11.1 & 2.84 & 92.1 \\
& Medium ($0.4$\,s $\leq$ RT60 $<1.0$\,s) & 8.0 & 8.3 & 2.61 & 88.7 \\
& High (RT60 $\geq1.0$\,s) & 5.4 & 5.7 & 2.40 & 86.4 \\
\midrule
\multirow{2}{*}{Speaker motion}
& Static speakers & 10.3 & 10.8 & 2.79 & 90.2 \\
& Moving speakers & 3.7 & 4.5 & 2.42 & 84.1 \\
\bottomrule
\end{tabular}}
\vspace{-0.5em}
\end{table}

\textbf{Delay estimation and interference strength.}
Pseudo-target construction relies on estimating the delay between the excitation speech and each RM signal. Large delay-estimation errors degrade the resulting spatial supervision: SI-SDR falls from 10.7\,dB for errors below 0.625\,ms to 7.6\,dB for larger errors, with corresponding decreases in PESQ and STOI. Strong interference can corrupt the cross-correlation used for delay estimation. Consistent with this sensitivity, SI-SDR decreases from 15.2\,dB for SIR at or above 0\,dB to 7.0\,dB for SIR below 0\,dB. Delay errors and interference strength are related factors; these results characterize performance under the reported conditions rather than isolating their independent effects.

\textbf{Reverberation time.} The low-, medium-, and high-RT60 bins account for approximately 90\%, 8\%, and 2\% of the test samples, respectively.
Performance decreases as reverberation becomes stronger. SI-SDR falls from 10.6\,dB for RT60 below 0.4\,s to 8.0\,dB for the intermediate range and 5.4\,dB for RT60 at or above 1.0\,s. PESQ and STOI exhibit the same trend. These results identify long reverberation times as a challenging condition for the current pseudo-target construction and training setup, complementing the RIR-modeling analysis in Appendix~\ref{app:realman}.

\textbf{Speaker motion.}
The model trained on simulated static scenes exhibits a substantial performance drop when evaluated on moving-speaker scenes generated with gpuRIR: SI-SDR decreases from 10.3 to 3.7\,dB, PESQ from 2.79 to 2.42, and STOI from 90.2 to 84.1\%. This highlights the mismatch between static training conditions and time-varying acoustic paths. As shown in Appendix~\ref{app:realman}, incorporating real-recorded data that include moving speakers during training yields improvements over the four-channel baseline in moving-speaker recordings. Together, the two evaluations highlight the sensitivity to speaker motion and the importance of representing dynamic acoustic conditions during training.

\section{Real-time and array-agnostic variants for virtual microphone-based speech enhancement}\label{app:rt_aga}
\subsection{Real-time model}

Tables \ref{tab:virtual_flow_rt} and \ref{tab:virtual_flow_aa} evaluate the proposed Virtual-Flow in the VM-SE setting with its practicality under real-time constraints and its generalization to array-agnostic scenarios. We enforce causality throughout the system by replacing the convolution and layer-normalization operations with their causal counterparts for the real-time model. In addition, applying iSTFT to convert an STFT-based model output from the spectrogram domain back to the time domain introduces additional latency proportional to the windowing and overlap. To prevent this latency increase, we directly convert the output of the pre-trained lightweight Spatial-Magnifier from the STFT domain to the mel-scale, without performing iSTFT. Finally, in the real-time model, we add a skip connection from the lightweight Spatial-Magnifier output to the STFT of the PeriodWave-VME output, and construct the final neural-VME output in the STFT domain.

Table \ref{tab:virtual_flow_rt} focuses on real-time VM-SE. TinyGRU \cite{pandey2025ultra} is adopted for the baseline MC-SE model, which has ultra-low computational complexity. The real-time lightweight Spatial-Magnifier (LW SM-RT) increases VM-SE performance. Our Virtual-Flow-RT further improves VM-SE performance while satisfying causality and maintaining a 16\,ms algorithmic latency, demonstrating that VM-SE can remain effective even under strict streaming constraints. The resulting RTF is close to 1 (0.9424), implying limited computational headroom and motivating future optimization.

\begin{table*}[h]
\vspace{-0.5em}
\caption{VM-SE performance of real-time Virtual-Flow}
\label{tab:virtual_flow_rt}
\vspace{-0.5em}
\centering
\setlength{\tabcolsep}{1pt}
\renewcommand{\arraystretch}{1.1}
\scalebox{0.78}{
\begin{tabular}{clcccc cc}
\toprule
\multirow{2}{*}{Scenario} & \multirow{2}{*}{Method} & \multicolumn{4}{c}{VM-SE Performance} & \multicolumn{2}{c}{Efficiency} \\
\cmidrule(lr){3-6} \cmidrule(lr){7-8}
& & SI-SDR$\uparrow$ & SNR$\uparrow$ & PESQ$\uparrow$ & STOI$\uparrow$ & Latency & RTF \\
\midrule
& TinyGRU (4ch) & -4.2 & 1.7 & 1.57 & 59.8 & 16\,ms & \textbf{0.0478} \\ \cline{2-8}
& \quad + LW SM-RT & -2.3 & 2.4 & 1.69 & 63.3 & 16\,ms & 0.7188 \\
\rowcolor{gray!12} \cellcolor{white} \multirow{-2}{*}{Real-time} 
& \quad + Virtual-Flow-RT & \textbf{-1.7} & \textbf{2.7} & \textbf{1.75} & \textbf{64.2} & 16\,ms & 0.9424 \\ \cline{2-8}
& TinyGRU (16ch) & -2.0 & 2.4 & 1.80 & 67.5 & 16\,ms & 0.0478 \\
\bottomrule
\end{tabular}
}
\vspace{-1.0em}
\end{table*}

\subsection{Array-agnostic model}
Array-agnostic processing refers to the setting where the number and spatial configuration of microphones in the array vary across examples. To support array-agnostic processing, we propose an array transform strategy. Specifically, except for the reference microphone, we replace all remaining RM signals with VM signals. This transformation is intended to align the input distribution between training and inference, since RM microphone locations can follow different and potentially mismatched distributions across the two stages. In this setting, the lightweight Spatial-Magnifier adopts the transform-average-concatenate (TAC) module \cite{luo2020end}, which is widely used for array-agnostic processing. Concretely, the first up-block is processed in parallel for each microphone, and the resulting per-microphone features are averaged over the microphone-channel dimension, as in VarArray \cite{yoshioka2022vararray}. The rest of the pipeline remains identical to the original design. Finally, PeriodWave-VME is configured to output $M-1$ channels. Since $M_r$ varies across inputs, we pad the RM signals by repeating the reference-channel RM noisy signal $M - M_r - 1$ times. The subsequent steps follow the same procedure as in the baseline method.

Then, Table \ref{tab:virtual_flow_aa} reports array-agnostic VM-SE results. SpatialNet (Ran) exhibits very low performance, whereas applying the proposed array transform strategy with array-agnostic lightweight Spatial-Magnifier LW SM-AA and Virtual-Flow-AA (array-agnostic) yields substantial improvements across SI-SDR, SNR, PESQ, and STOI. In particular, Virtual-Flow-AA achieves the best perceptual and intelligibility scores (PESQ $2.01$, STOI $66.5$), verifying that our approach generalizes to varying microphone counts and geometries. Compared with DeFTAN-AA (Ran) \cite{lee2024deftan}, which is the state-of-the-art array geometry-agnostic speech enhancement model, our method attains higher speech quality and intelligibility with a favorable performance--complexity trade-off, supporting its applicability to practical array-agnostic scenarios.

\begin{table*}[h]
\vspace{-0.5em}
\caption{VM-SE performance of array-agnostic Virtual-Flow}
\label{tab:virtual_flow_aa}
\vspace{-0.5em}
\centering
\setlength{\tabcolsep}{1pt}
\renewcommand{\arraystretch}{1.1}
\scalebox{0.78}{
\begin{tabular}{clcccc cc}
\toprule
\multirow{2}{*}{Scenario} & \multirow{2}{*}{Method} & \multicolumn{4}{c}{VM-SE Performance} & \multicolumn{2}{c}{Efficiency} \\
\cmidrule(lr){3-6} \cmidrule(lr){7-8}
& & SI-SDR$\uparrow$ & SNR$\uparrow$ & PESQ$\uparrow$ & STOI$\uparrow$ & Param. & MAC/s \\
\midrule
& SpatialNet (Ran) & -11.7 & 0.2 & 1.34 & 43.9 & 1.6\,M & 23.2\,G \\ \cline{2-8}
& \quad + LW SM-AA & -4.7 & 3.8 & 1.90 & 63.7 & \textbf{+0.25\,M} & \textbf{+2.0\,G} \\
\rowcolor{gray!12} \cellcolor{white} \multirow{-2}{*}{Agnostic} 
& \quad + Virtual-Flow-AA & \textbf{-3.7} & \textbf{4.8} & \textbf{2.01} & \textbf{66.5} & +13.1\,M & +7.8\,G \\
& DeFTAN-AA (Ran) & -8.4 & 1.2 & 1.48 & 52.3 & 2.2\,M & 35.7\,G \\
\bottomrule
\end{tabular}
}
\vspace{-1.0em}
\end{table*}

\section{Estimated VM location plot}\label{app:vm_location}
\begin{figure}[!h]
\vspace{-1em}
\centering
\includegraphics[width=3.5in]{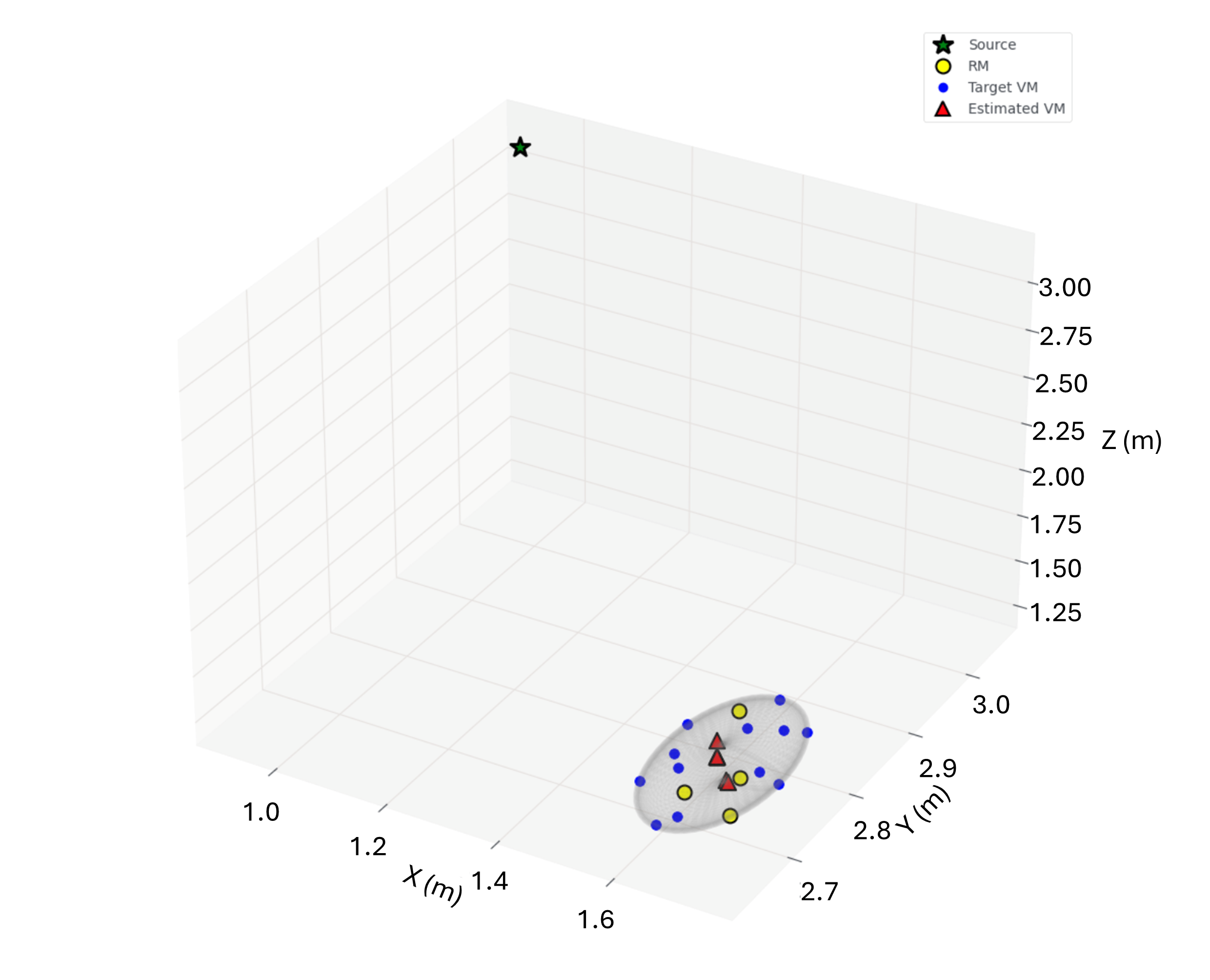}
\vspace{-0.5em}
\caption{Estimated VM location plot}
\label{fig:location}
\vspace{-1em}
\end{figure}

Figure~\ref{fig:location} provides a qualitative visualization of the estimated VM locations produced by the proposed Virtual-Flow trained in the semi-supervised setting. Although the model never observes VM noisy signals during training, the estimated VM channels (red triangles) are arranged with the target VM geometry (blue dots), forming a random spatial distribution in the virtual sphere implied by the array configuration. 
This observation suggests that the model effectively injects extra spatial samples into the input space, resembling the availability of additional microphones, even under the constraint that VM signals are inaccessible.

\section{Spectrogram analysis}
\subsection{Estimated VM spectrogram comparison}\label{app:VME_spec}
\begin{figure*}[!h]
\centering
\includegraphics[width=5.5in]{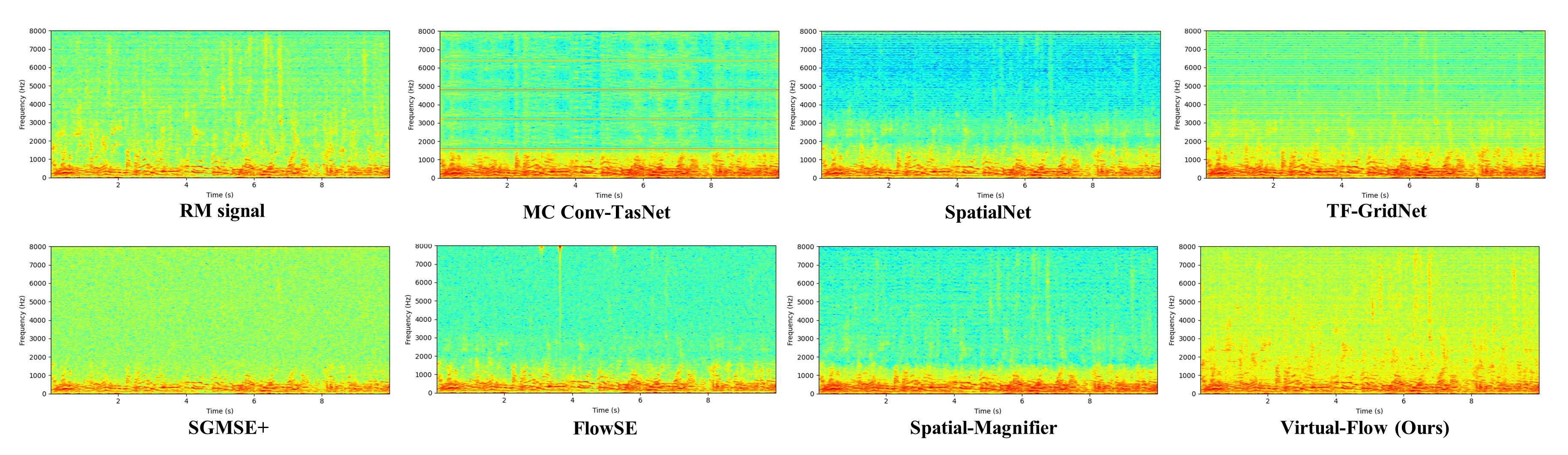}
\vspace{-1.5em}
\caption{Estimated VM spectrograms compared with noisy reference VM signals.}
\label{fig:spec}
\vspace{-1em}
\end{figure*}

The reference in Figure~\ref{fig:spec} is a noisy VM signal, rather than the clean target speech used to supervise downstream enhancement. Neural-VME is evaluated against the spectral structure of these noisy microphone observations; the final enhanced clean-reference estimate is a separate output of the VM-BF pipeline. The comparison therefore examines whether generated channels preserve microphone-like spectrotemporal structure, rather than whether they directly reproduce clean target speech. This qualitative evidence should be considered jointly with the MVDR and co-directional results, rather than as a standalone proof that all gains are spatial.

From the spectrogram comparisons (Figure \ref{fig:spec}), the proposed Virtual-Flow exhibits the most faithful reconstruction of the target spectral structure, particularly in the high-frequency bands. Several baseline estimates exhibit attenuated or oversmoothed spectral regions relative to the noisy VM reference, particularly in the upper bands, leading to a loss of fine spectral details that are important for spatial cue preservation and speech clarity. In contrast, Virtual-Flow better preserves the distribution of energy and temporal variations across frequencies, and its spectrogram remains visually the closest to the reference spectral pattern among the compared models. This qualitative evidence is consistent with our quantitative findings that emphasize the importance of high-frequency restoration for effective spatial enhancement and downstream performance.

\subsection{Spectrogram comparison between Spatial-Magnifier and Virtual-Flow}\label{app:spectrogram}
\begin{figure*}[!h]
\centering
\includegraphics[width=5.5in]{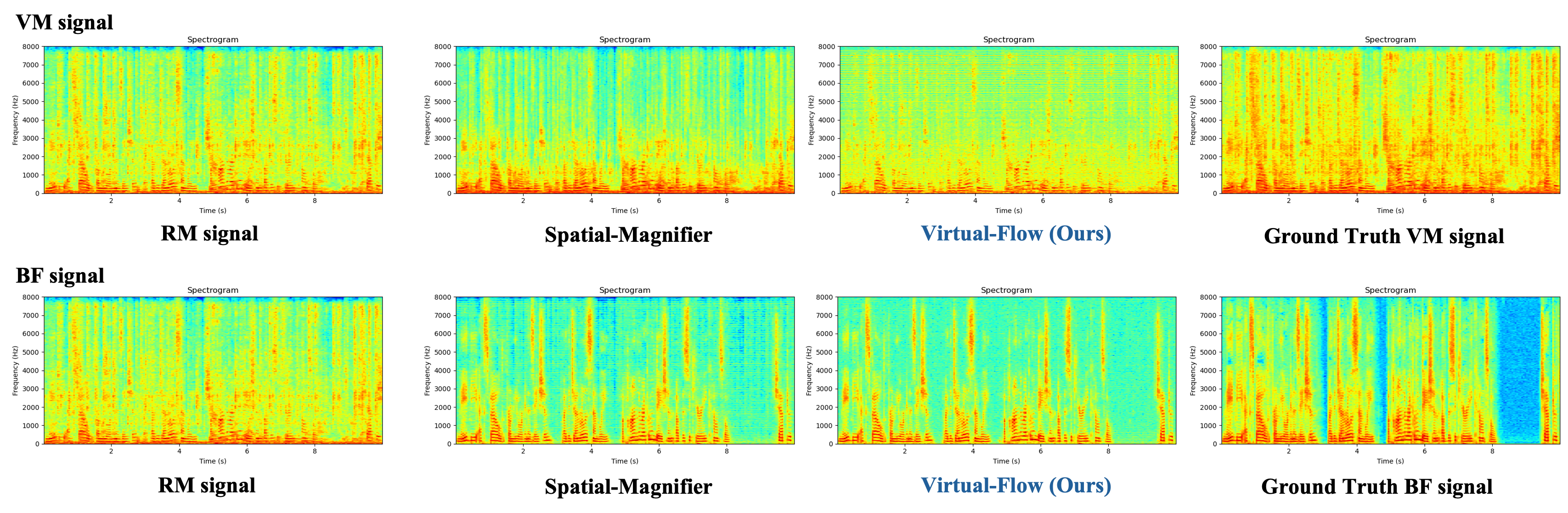}
\vspace{-1.5em}
\caption{Spectrogram comparison between Spatial-Magnifier and Virtual-Flow}
\label{fig:spec_comp}
\vspace{-1em}
\end{figure*}

Consistent with the quantitative results, the spectrogram comparison in Figure~\ref{fig:spec_comp} demonstrates that Virtual-Flow successfully captures the detailed speech distribution in high-frequency bands, whereas Spatial-Magnifier fails to reconstruct these critical components. This visual evidence provides a clear rationale for the performance trade-offs observed: while Virtual-Flow exhibits lower Neural-VME performance in time-domain metrics (e.g., SI-SDR, SNR), it significantly outperforms baselines in magnitude-based metrics such as M-STFT and LSD. The superior recovery of high-frequency magnitude information directly mitigates spatial aliasing, thereby enhancing overall spatial resolution. Furthermore, although the differences in the final VM-BF signals are subtle to the naked eye, a closer inspection reveals that Virtual-Flow more accurately restores high-frequency speech structures compared to Spatial-Magnifier, which is fundamental for achieving robust multichannel speech extraction.


\newpage
\input{checklist.tex}

\end{document}

%% file: checklist.tex
\section*{NeurIPS Paper Checklist}

\begin{enumerate}

\item {\bf Claims}
    \item[] Question: Do the main claims made in the abstract and introduction accurately reflect the paper's contributions and scope?
    \item[] Answer: \answerYes{} 
    \item[] Justification: Our key contribution is a semi-supervised framework for VM-BF and VM-SE that uses clean reference speech during training without ground-truth VM recordings, and requires only RM recordings at inference. By estimating a large number of VM signals (more than 10), we significantly leverage the inherent advantages of neural-VME. These contributions are explicitly highlighted in the abstract and introduction.

\item {\bf Limitations}
    \item[] Question: Does the paper discuss the limitations of the work performed by the authors?
    \item[] Answer: \answerYes{} 
    \item[] Justification: As explicitly stated in the conclusion, the proposed research has a few limitations. Reflow requires two training stages, and the computational cost remains challenging for low-power on-device deployment. Moving speakers are evaluated in Appendices~\ref{app:realman} and~\ref{app:sensitivity}: a model trained on simulated static scenes degrades on moving-speaker tests, while training on real-recorded data that include motion improves performance. Pseudo-target construction and phase estimation also remain approximate.

\item {\bf Theory assumptions and proofs}
    \item[] Question: For each theoretical result, does the paper provide the full set of assumptions and a complete (and correct) proof?
    \item[] Answer: \answerYes{} 
    \item[] Justification: Detailed descriptions of the neural-VME task and flow matching theory, along with comprehensive experimental results supporting our claims, are thoroughly presented in both the main text and the appendix.

    \item {\bf Experimental result reproducibility}
    \item[] Question: Does the paper fully disclose all the information needed to reproduce the main experimental results of the paper to the extent that it affects the main claims and/or conclusions of the paper (regardless of whether the code and data are provided or not)?
    \item[] Answer: \answerYes{} 
    \item[] Justification: We detail the experimental conditions, datasets, model parameters, and training procedures in the Experiments section.

\item {\bf Open access to data and code}
    \item[] Question: Does the paper provide open access to the data and code, with sufficient instructions to faithfully reproduce the main experimental results, as described in supplemental material?
    \item[] Answer: \answerNo{} 
    \item[] Justification: We cannot release the code due to internal company policies.

\item {\bf Experimental setting/details}
    \item[] Question: Does the paper specify all the training and test details (e.g., data splits, hyperparameters, how they were chosen, type of optimizer) necessary to understand the results?
    \item[] Answer: \answerYes{} 
    \item[] Justification: We detail the experimental settings, model parameters, and training procedures in the Experiments section.

\item {\bf Experiment statistical significance}
    \item[] Question: Does the paper report error bars suitably and correctly defined or other appropriate information about the statistical significance of the experiments?
    \item[] Answer: \answerYes{} 
    \item[] Justification: For the baseline MCWF comparison, supervised Spatial-Magnifier and semi-supervised Virtual-Flow are evaluated over three independent runs with different random seeds. The Experiments section defines the reported mean and standard deviation, and the corresponding existing tables report mean $\pm$ standard deviation. Uncertainty estimates are not reported for all other experimental conditions; no formal hypothesis test is claimed.

\item {\bf Experiments compute resources}
    \item[] Question: For each experiment, does the paper provide sufficient information on the computer resources (type of compute workers, memory, time of execution) needed to reproduce the experiments?
    \item[] Answer: \answerYes{} 
    \item[] Justification: We have specified the type and number of GPUs used for the experiments in the experimental setting section.
    
\item {\bf Code of ethics}
    \item[] Question: Does the research conducted in the paper conform, in every respect, with the NeurIPS Code of Ethics \url{https://neurips.cc/public/EthicsGuidelines}?
    \item[] Answer: \answerYes{} 
    \item[] Justification: we confirm that the research conducted in this paper strictly conforms to the NeurIPS Code of Ethics in every respect.

\item {\bf Broader impacts}
    \item[] Question: Does the paper discuss both potential positive societal impacts and negative societal impacts of the work performed?
    \item[] Answer: \answerYes{} 
    \item[] Justification: We have explicitly discussed the broader impacts of our proposed framework in the conclusion, emphasizing its potential to democratize high-quality spatial audio for physically constrained devices like AR glasses and hearing aids. Furthermore, we highlighted how our semi-supervised approach reduces the environmental and computational costs associated with large-scale data collection.
    
\item {\bf Safeguards}
    \item[] Question: Does the paper describe safeguards that have been put in place for responsible release of data or models that have a high risk for misuse (e.g., pre-trained language models, image generators, or scraped datasets)?
    \item[] Answer: \answerNA{} 
    \item[] Justification: This work is not related to high risk for misuse.

\item {\bf Licenses for existing assets}
    \item[] Question: Are the creators or original owners of assets (e.g., code, data, models), used in the paper, properly credited and are the license and terms of use explicitly mentioned and properly respected?
    \item[] Answer: \answerYes{} 
    \item[] Justification: We have cited all data used for training and existing model structures, and these are all open.

\item {\bf New assets}
    \item[] Question: Are new assets introduced in the paper well documented and is the documentation provided alongside the assets?
    \item[] Answer: \answerNA{} 
    \item[] Justification: This work is not related to new assets

\item {\bf Crowdsourcing and research with human subjects}
    \item[] Question: For crowdsourcing experiments and research with human subjects, does the paper include the full text of instructions given to participants and screenshots, if applicable, as well as details about compensation (if any)? 
    \item[] Answer: \answerNA{} 
    \item[] Justification: This work is not related to crowdsourcing experiments.

\item {\bf Institutional review board (IRB) approvals or equivalent for research with human subjects}
    \item[] Question: Does the paper describe potential risks incurred by study participants, whether such risks were disclosed to the subjects, and whether Institutional Review Board (IRB) approvals (or an equivalent approval/review based on the requirements of your country or institution) were obtained?
    \item[] Answer: \answerNA{} 
    \item[] Justification: This work is not related to IRB.

\item {\bf Declaration of LLM usage}
    \item[] Question: Does the paper describe the usage of LLMs if it is an important, original, or non-standard component of the core methods in this research? Note that if the LLM is used only for writing, editing, or formatting purposes and does \emph{not} impact the core methodology, scientific rigor, or originality of the research, declaration is not required.
    \item[] Answer: \answerNA{} 
    \item[] Justification: In this research, LLMs were used strictly for simple grammar correction and polishing, and are completely unrelated to the core scientific content of the study.

\end{enumerate}